\documentclass{aa}

\newcommand{\magpt}[2]{\mbox{$\rm #1\hspace{-0.25em}\stackrel{m}{.}
      \hspace{-1.0mm}#2$}}                             % magnitude \magpt {}{}
\def\bsec{\hbox{$.\!\!{\arcsec}$}}

\newcommand\RA[3]{#1$^{\rm h}$#2$^{\rm m}$#3$^{\rm s}$}
\newcommand\DEC[3]{#1$^{\circ}$#2\arcmin#3\arcsec}

\newcommand\teff{$T_{\rm eff}$}

\newcommand\logg{$\log~g$}

\newcommand\ebv{$ {\rm E_{B-V}}$}
\newcommand{\Msolar}{\mbox{\,$\rm M_{\odot}$}}        % solar mass
\newcommand{\Rsolar}{\mbox{\,$\rm R_{\odot}$}}        % solar mass
\newcommand{\dgpt}[2]{\mbox{\rm #1\hspace{-0.10em}$\stackrel{\circ}{.}$
      \hspace{-1.0mm}#2}}                             % \dgpt {}{}
\newcommand{\fbsdrei}{{\it Proceedings of the
Third Conference on Faint Blue Stars, {\rm eds. A.G.D. Philip,  
J. Liebert and R.A. Saffer, Schenectady: L.Davis Press}}}
\newcommand{\wdt}{{\it White Dwarfs, {\rm eds. G. Vauclair, E. Sion, 
Dordrecht, Kluwer}}}
\begin{document}
%\thesaurus{05 ()}
\title{Resolving subdwarf B stars in binaries by HST imaging
\thanks{Based on observations with the NASA/ESA 
{\it Hubble Space Telescope} obtained at the Space Telescope Science 
Institute, which is operated by the Association of Universities for 
Research in Astronomy, Inc., under NASA contract NAS 5-26555}
\thanks{Based on observations collected at the German-Spanish Astronomical 
Center (DSAZ), Calar Alto, operated by the Max-Planck-Institut f\"ur 
Astronomie Heidelberg jointly with the Spanish National Commission for 
Astronomy} 
\thanks{Based on data obtained at ESO (ESO proposal No. 
58.D-0478, 65.H-0253(A))}
\thanks{Some observations reported here were obtained at the MMT Observatory, a 
joint facility of the University of Arizona and the Smithsonian Institution.}}

\author{U. Heber\inst{1} \and S. Moehler\inst{1} \and R. Napiwotzki\inst{1}
P. Thejll\inst{2} \and E.M. Green\inst{3}} 
\offprints{U. Heber}
\institute {Dr. Remeis-Sternwarte, Astronomisches Institut der Universit\"at
Erlangen-N\"urnberg, Sternwartstr. 7, 96049 Bamberg, Germany
\and
Solar-Terrestrial Physics Division, Danish Meteorological Institute,
Lyngbyvej 100, DK-2100 Copenhagen O, Denmark
\and
Steward Observatory, University of Arizona, Tucson, AZ~85721, USA}
%\date{Received,  accepted }
\date{}
\abstract{
The origin of subluminous B stars is still an unsolved problem in stellar 
evolution. Single star as well as close binary evolution scenarios have been 
invoked but until now have met with little success. We have carried out 
a small survey of spectroscopic binary candidates 
(19 systems consisting of an sdB 
star and late type companion) with the Planetary Camera of the WFPC2 onboard 
Hubble Space Telescope to test these scenarios. 
Monte Carlo simulations 
indicate that by imaging the programme stars in the R-band 
about one third of the sample (6--7 stars) should be resolved at a limiting 
angular resolution 
of 0\bsec1 if they have linear separations like main sequence stars 
(``single star evolution''). 
None should be resolvable if all systems were produced by close binary 
evolution. In addition we expect three triple systems to 
be present in our sample. Most of these, if not all, should be resolvable.   
Components were resolved in 6 systems with separations between 0\bsec2 and 
4\bsec5. However, only in the two systems TON~139 and PG~1718$+$519 
(separations 0\bsec32 and 0\bsec24, respectively)
do the magnitudes of the resolved components match the expectations from 
the deconvolution of the spectral energy distribution.
These two stars could be physical binaries whereas in the other cases 
the nearby star may be a chance projection or a third component.
Radial velocity measurements indicate that the resolved system 
TON~139 is a triple system, with the sdB having a close 
companion that does not contribute detectably to the integrated
light of the system. Radial velocity information for the 
second resolved system, PG~1718$+$519, is insufficient. Assuming that it is not a 
triple system, it would be the only resolved system in our sample.
Accordingly the success rate would be only 5\% which is clearly {\it below} 
the prediction for single star 
evolution. 
%but {\it above} that for close binary evolution.  
%%If the latter scenario were valid for all sdB stars, the two resolved systems
%would have to be triple system. 
We conclude that the distribution of separations of sdB binaries deviates
strongly from that of normal stars.
Our results add further evidence that close binary evolution is
fundamental for the evolution of sdB stars.
}

\maketitle

\section{Introduction\label{intro}} 
 
Subluminous B (sdB) stars dominate the populations of faint blue 
stars of our own Galaxy and are found in both the disk (field sdBs) and 
globular clusters (Moehler et al. \cite{mohe97}).
Observations of elliptical galaxies with the Ultraviolet Imaging Telescope
(Brown et al.\ \cite{brfe97})
and the Hubble Space Telescope (Brown et al.\ \cite{brbo00}) 
have shown  that these stars are sufficiently common to be the dominant  
source for the ``UV upturn phenomenon'' 
observed in elliptical galaxies and galaxy bulges (see also Greggio \& Renzini
\cite{grre90}, \cite{grre99}). 
Their space distribution and kinematical properties indicate that the field 
stars belong to the intermediate to old disk population (de Boer et al. 
\cite{deag97}; Altmann \& de Boer \cite{alde00}). 

However, important questions remain 
concerning their formation process and the appropriate evolutionary timescales.
This is a major drawback for the calibration of the observed ultraviolet 
upturn in elliptical galaxies as an age indicator.  

It is now generally accepted that the sdB 
stars can be identified with 
models for Extreme Horizontal Branch (EHB) stars burning He in their core, 
but with a very tiny ($<$2\% by mass) inert hydrogen envelope (Heber 
\cite{hebe86}; Saffer et al.\ \cite{sabe94}).
An EHB star bears great resemblance to a helium main-sequence star 
of half a solar mass and its further evolution should proceed similarly 
(i.e. directly to the white dwarf graveyard) 
as confirmed by evolutionary calculations (Dorman et al. \cite{doro93}).

How stars evolve to the EHB configuration is controversial. 
The problem is how the mass loss mechanism in the
progenitor manages
to remove all but a tiny fraction of the hydrogen envelope at {\em
precisely} the same time as the He core has attained the minimum
mass ($\approx0.5$M$_\odot$) required for the He flash. 

Both non-interacting (scenario i), and interacting (scenarios ii and iii)
 evolutionary scenarios have been proposed to explain the origin of the
sdB stars (see Bailyn et al. \cite{basa92}). 

\noindent (i) Enhanced mass loss on the red giant branch
(RGB) before or during the core helium flash
may remove almost the entire hydrogen-rich envelope. This is usually 
modelled by increasing the $\eta$ factor in the Reimers (\cite{reim75}) formula
to estimate mass loss rates for RGB stars. It has been conjectured that the 
mass loss rates increase with increasing metallicity, implying 
that metal rich populations should produce more sdB stars 
than metal poor ones. 
Birthrate estimates for sdB stars 
indicate that only 2\% (Heber \cite{hebe86}) or even less (0.25\% to 1\%, 
Saffer \& Liebert \cite{sali95}) of the RGB stars
need to experience such enhanced mass loss. Evidence that this is
possible comes from the existence of RR Lyrae stars of population I which
must also have lost half of their mass during evolution. In both cases
the physical reason for such strong mass loss is not yet understood. 

\noindent (ii) Mengel et al. (\cite{meno76}) suggest that sdBs could be formed
from binaries in which mass transfer starts
on the red giant branch and results in a reduction of the hydrogen envelope
prior to the helium core flash. Hence all sdBs star are predicted to be
found in close binary systems. 

\noindent (iii) An alternative scenario was proposed by Iben
(\cite{iben90}), who pointed out that sdBs can be formed from mergers of
helium white dwarf binary systems. Iben \& Tutukov (\cite{ibtu92})
estimate that 80\% of the sdBs could have been formed by mergers. Hence the
frequency of sdBs still being in binaries should be at most 20\%. 

Several dozens of objects with composite spectra consisting of an sdB and a
dwarf G-K star have been discovered (e.g. Ferguson et al.
\cite{fegr84}; Theissen et al. \cite{thmo93},
\cite{thmo95}; Allard et al. \cite{alwe94}) which implies that the binary
frequency of sdBs is 50\% or more (Allard et al. \cite{alwe94}). 
The observed large binary frequency rules out the merger scenario (iii) and we
are left with scenarios (i) and (ii), i.e. either the sdB binaries are
mostly wide systems that did not interact so that the sdB
precursors have evolved independently from the companion (i), or they are
close systems formed by interaction of the sdB precursor with the companion
star (mass exchange, ii). 

The high spatial resolution of the {\it Planetary Camera} (PC) on board the 
{\it Hubble Space Telescope} (HST) allows to perform a crucial test. 
As we will show in this 
paper, it should be possible to resolve a significant fraction of the known 
composite spectrum systems containing an sdB star if scenario (i) is correct, 
i.e. if the systems have a 
distribution of separations like normal main sequence binaries 
(Duquennoy \& Mayor \cite{duma91}).
The interacting scenario (ii), however, predicts that all 
sdB stars reside in short period (P$\le$100d) binaries and consequently none 
of the systems should be resolvable even with the PC. 
In order to measure their distribution of separations we have 
imaged 23 sdB binary candidates with the PC by taking advantage of 
the snap shot mode of HST observations.

\section{Observations and Data Analysis\label{obs}}

\begin{table*}
\caption[]
{Programme stars: coordinates, observation dates, and exposure times for the 
WFPC2 and references for the spectroscopic classification 
observations\label{tab_targ}}
\begin{tabular}{lrrrrrrl}
\hline
  star & $\alpha_{1950}$ & $\delta_{1950}$ & l & b & obs. & exp. & 
reference\\
       &                 &                 & & & date & time & \\
       &                 &                 &  &  &  & [s] & \\
\hline
PB~6107 & \RA{00}{39}{31} & \DEC{+04}{53}{17} & \dgpt{118}{59} &
\dgpt{$-$57}{64} & 990627 & 3.5 & Moehler et al. (\cite{mori90})\\
PHL~1079 & \RA{01}{35}{48} & \DEC{+03}{23}{00} & \dgpt{144}{96} & 
\dgpt{$-$57}{22} & 981204 & 4 & Theissen et al. (\cite{thmo95})\\
HE~0430$-$2457 & \RA{04}{30}{59} & \DEC{$-$24}{57}{37} & \dgpt{223}{49} 
& \dgpt{$-$40}{55} & 980417 & 8 & this paper\\
PG~0749$+$658 & \RA{07}{49}{39} & \DEC{+65}{50}{13} & \dgpt{150}{44} & 
\dgpt{$+$30}{99} & 990329 & 1.8 & Saffer (\cite{saff91})\\
PG~0942$+$461 & \RA{09}{42}{02}& \DEC{+46}{08}{38} & \dgpt{173}{11} & 
\dgpt{$+$48}{89} & 980530 & 10 & Heber et al. (\cite{hjw91})\\ 
TON~1281 & \RA{10}{40}{57} & \DEC{+23}{24}{55} & \dgpt{213}{62} & 
\dgpt{$+$60}{89} & 990623 & 5 & Jeffery \& Pollacco (\cite{jepo98})\\
TON~139 & \RA{12}{53}{39} & \DEC{+28}{23}{31} & \dgpt{77}{21} & 
\dgpt{$+$88}{57} & 980103 & 1.8 & Green (\cite{gree97})\\
PG~1309$-$078 & \RA{13}{09}{09} & \DEC{$-$07}{49}{18} & \dgpt{311}{60} & 
\dgpt{$+$54}{44} & 980505 & 8 & Ferguson et al. (\cite{fegr84})\\
PG~1421$+$345 & \RA{14}{21}{29} & \DEC{+34}{27}{53} & \dgpt{58}{36} & 
\dgpt{$+$69}{01} & 990605 & 14 & Ferguson et al. (\cite{fegr84})\\
PG~1449$+$653 & \RA{14}{49}{42} & \DEC{+65}{17}{58} & \dgpt{104}{84} & 
\dgpt{$+$47}{63} & 990619 & 7 & Moehler et al. (\cite{mori90}) \\
PG~1511$+$624 & \RA{15}{11}{25} & \DEC{+62}{21}{00}& \dgpt{99}{21} & 
\dgpt{$+$47}{96} & 990513 & 14 & Moehler et al. (\cite{mori90})\\
PG~1601$+$145 & \RA{16}{01}{47} & \DEC{$+$14}{32}{58} & \dgpt{27}{15} & 
\dgpt{$+$43}{51} & 000613 & 12 & Ferguson et al. (\cite{fegr84})\\
PG~1636$+$104 & \RA{16}{36}{40} & \DEC{$+$10}{24}{54} & \dgpt{27}{00} & 
\dgpt{$+$34}{04} & 000612 & 8 & Ferguson et al. (\cite{fegr84}) \\
TON~264 & \RA{16}{47}{05} & \DEC{+25}{15}{13} & \dgpt{45}{16} & 
\dgpt{$+$37}{12} & 990529 & 10 & Theissen et al. (\cite{thmo93})\\
PG~1656$+$213 & \RA{16}{56}{12} & \DEC{+21}{15}{05}& \dgpt{41}{25} & 
\dgpt{$+$33}{90}& 980301 & 12 & Ferguson et al. (\cite{fegr84}) \\
PG~1718$+$519 & \RA{17}{18}{35} & \DEC{+51}{55}{05}& \dgpt{79}{00} & 
\dgpt{$+$34}{94} & 990427 & 7 & Theissen et al. (\cite{thmo95})\\
PG~2148$+$095 & \RA{21}{48}{41} & \DEC{+09}{30}{39} & \dgpt{66}{78} & 
\dgpt{$-$32}{84} & 990411 & 4 & this paper\\
HE~2213$-$2212 & \RA{22}{13}{38} & \DEC{$-$22}{12}{26} & \dgpt{32}{63} & 
\dgpt{$-$54}{50} & 981207 & 8 & this paper\\ 
BD~$-$7$^\circ$5977 & \RA{23}{15}{12} & \DEC{$-$06}{44}{56} & \dgpt{71}{55} & 
\dgpt{$-$59}{65} & 981125 & 0.3 & Viton et al. (\cite{vi91})\\
\hline
\multicolumn{8}{c}{stars without spectroscopic evidence for a cool 
companion}\\
\hline
PG~0105$+$276 & \RA{01}{05}{32} & \DEC{+27}{36}{53} & \dgpt{127}{46} & 
\dgpt{$-$34}{84} & 980226 & 14 & this paper, new type: He-sdO\\
PG~1558$-$007 & \RA{15}{58}{39} & \DEC{$-$00}{43}{26} & \dgpt{9}{34} & 
\dgpt{$+$36}{51} & 990424 & 7 & this paper\\
KPD~2215$+$5037 & \RA{22}{15}{25} & \DEC{+50}{37}{48} & \dgpt{99}{71} & 
\dgpt{$-$4}{91} & 961213 & 7 & this paper\\
PG~2259$+$134 & \RA{22}{59}{16} &\DEC{$+$13}{22}{31} & \dgpt{86}{36} & 
\dgpt{$-$41}{31} & 000615 & 10 & Theissen et al. (\cite{thmo93}), this paper \\
\hline
\end{tabular}
\end{table*}

\subsection{Target selection and optical spectroscopy\label{targets}}

For the snapshot observations a target list of fifty of the brightest sdB
star binary candidates 
was extracted from an updated version of the Kilkenny et al.
(\cite{kilk88}) catalogue, supplemented by 
two stars which we discovered in
the course of follow-up spectroscopy of hot stars from the Hamburg-ESO
survey (see Edelmann et al. \cite{edel01}). 23 stars from this target list
were actually observed with the {\em Wide Field Planetary Camera 2} (WFPC2)
onboard the HST during our 
snapshot project, i.e. they were scheduled for observation to fill small
gaps in the HST schedule. All stars have published photometry (see
Tables~\ref{medium} and \ref{broad}), but only 16 have published optical
spectroscopy. Therefore additional spectra were obtained at the Calar Alto
and ESO observatories (see Appendix~\ref{app_spec} for details and plots of
the spectra in Fig.~\ref{bin_spec} and Fig.~\ref{sing_spec}). As can be seen
from Fig.~\ref{bin_spec} spectral features (\ion{Ca}{i}, \ion{Ca}{ii},
\ion{Mg}{i} and/or \ion{Fe}{i}) indicative of a cool star are clearly
present in the spectra of PG~1309$-$078, PG~0942$+$461, HE~0430$-$2457,
HE~2213$-$2212, and PG~2148$+$095 in addition to the Balmer and helium
lines of the sdB. Hence these objects are spectroscopic binaries consisting
of an sdB star and a cool companion. PG~0942$+$461 has already been
observed by Mitchell (\cite{mitc98}), who, however, did not note the binary
nature of the star. We do not find any evidence for a cool companion in the
spectra of the sdB stars PG~1558$-$087 and KPD~2215$+$5037 
(see Fig.~\ref{sing_spec}).
We also re-analysed a published spectrum of PG~2259$+$134 (Theissen et al.,
\cite{thmo93}) and do not find any spectroscopic evidence for a 
cool companion. 
PG~0105$+$276 turns out to be not an sdB star but a helium-rich sdO star and 
does not show any spectroscopic evidence for a cool companion. 
Therefore our sample consists of 19 composite spectrum objects plus four stars 
showing only photometric evidence for a companion. One of these four stars 
(PG~0105$+$276) also does not
belong to the programme sample since it is an sdO star.

\subsection{WFPC2 data\label{wfpc2}}

We observed the candidate binary systems with the PC chip of the WFPC2. If
the cool companion is a main sequence star, both components should be of
comparable brightness in the $R$ band and we therefore used the $F675W$ filter
of the WFPC2. We obtained four observations of each target, which were offset
relative to the first one by  ($-$11,$-$5.5), ($-$16.5,$-$16.5),
($-$5.5,$-$11) pixels. We first rebinned the data linearly to a step size
of 0.5 pixels and then aligned them according to the offset pattern
mentioned above. We then determined the median value of the four aligned
images to avoid cosmic ray hits and hot pixels and used these
median-averaged images for visual inspection. All flux measurements are
performed on manually cleaned average images to ensure proper flux
conservation. 

\begin{figure*}
\vspace{11cm}
\includegraphics{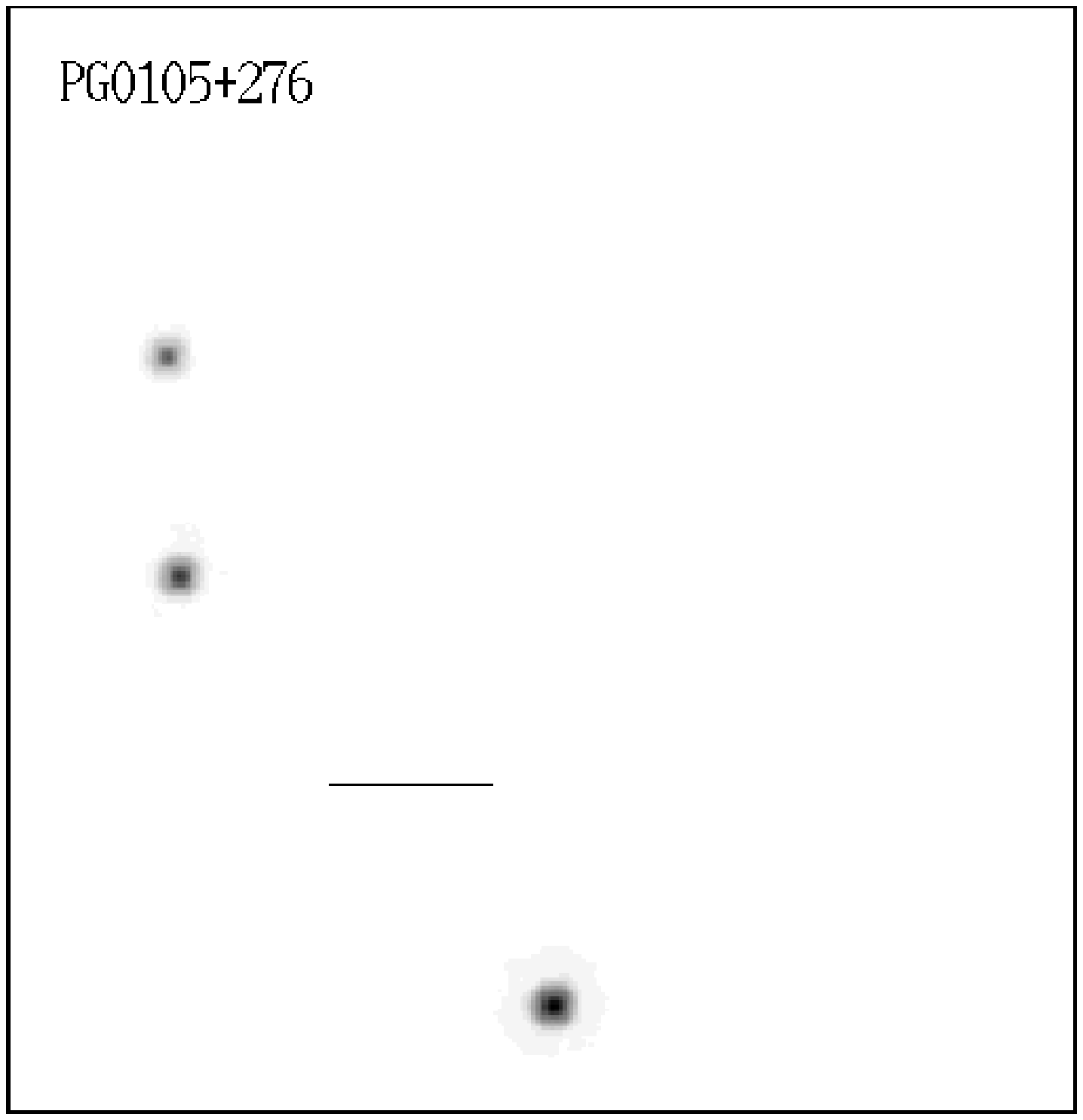}
\includegraphics{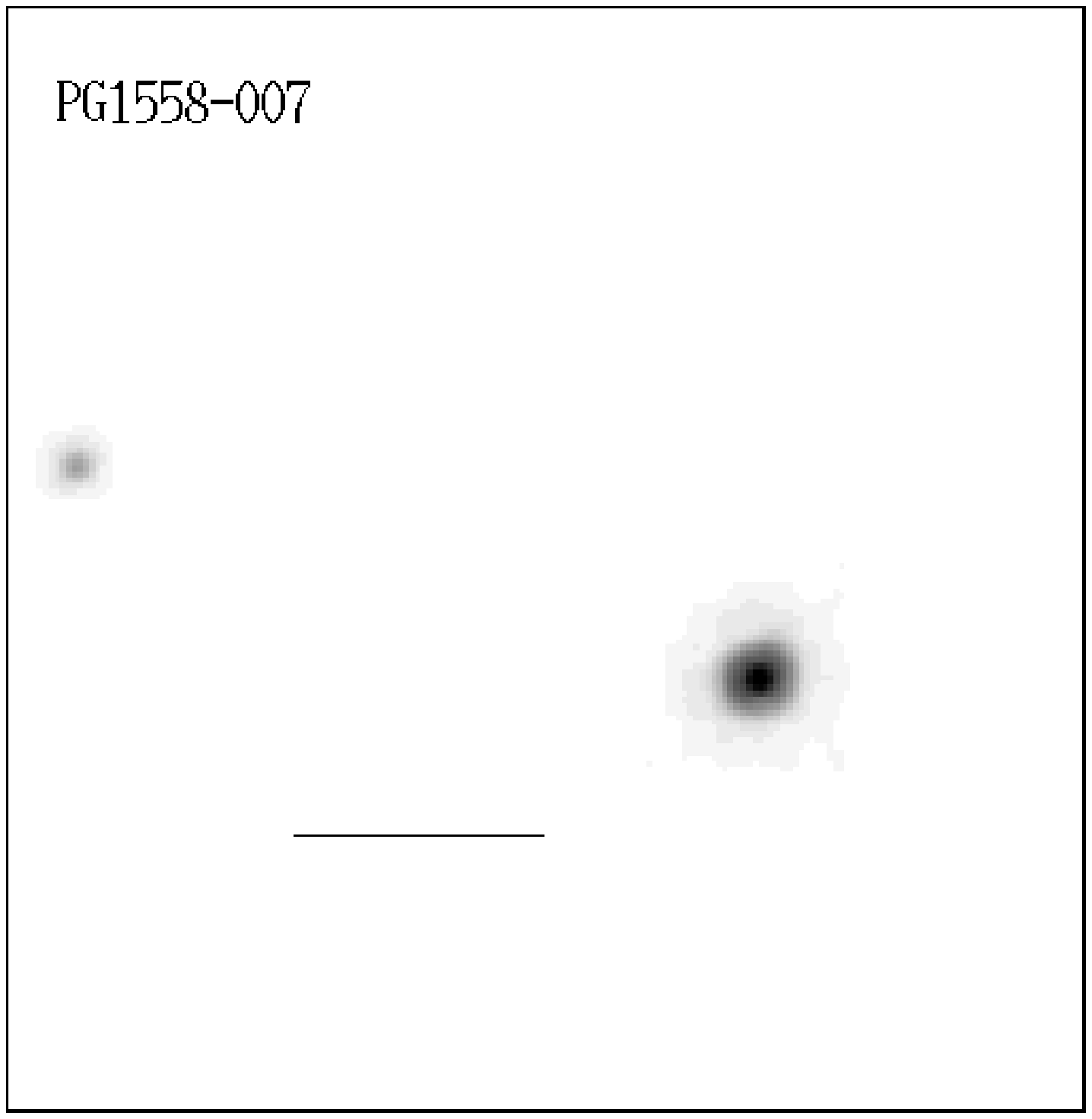}
\includegraphics{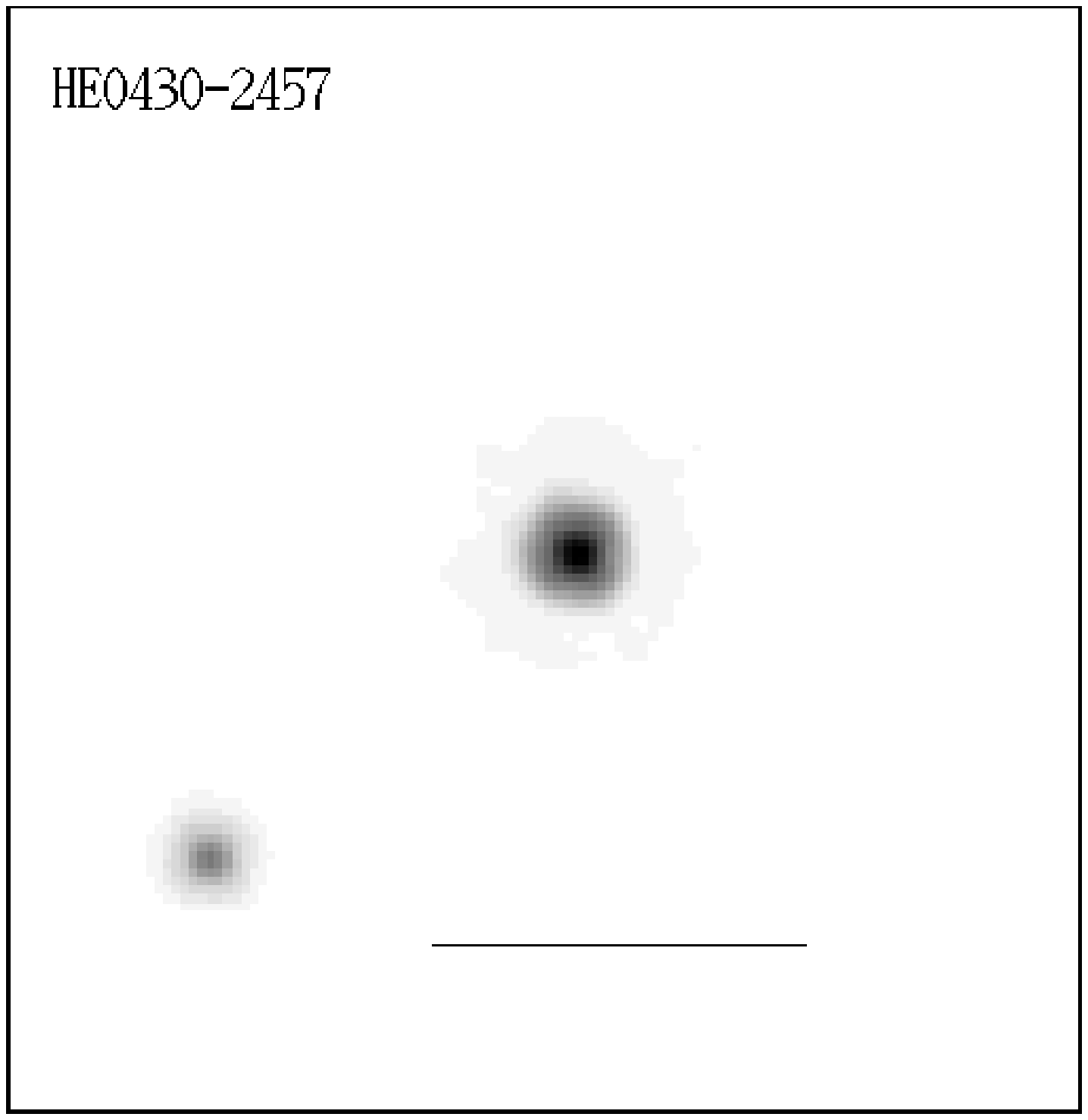}
\includegraphics{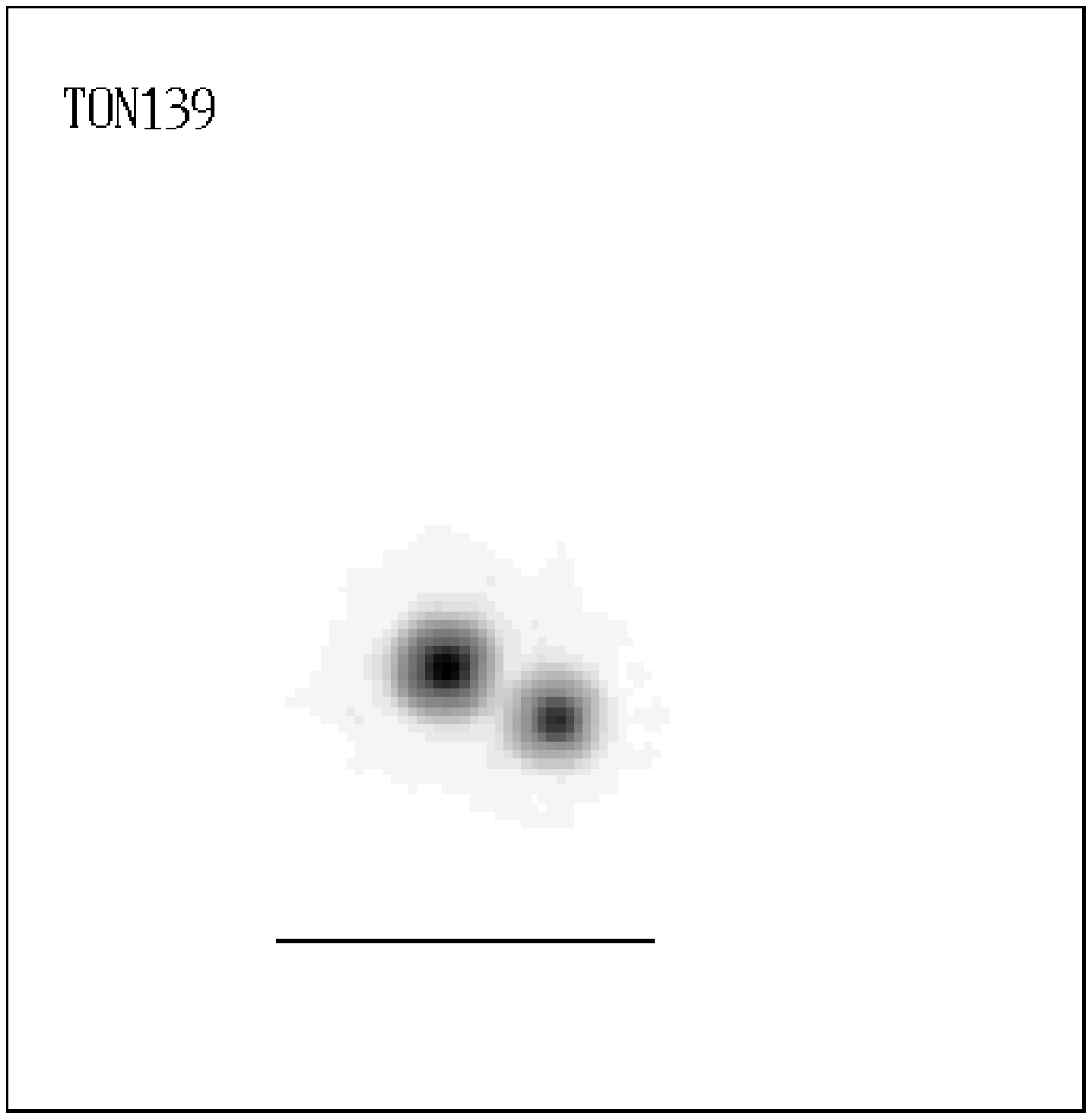}
\includegraphics{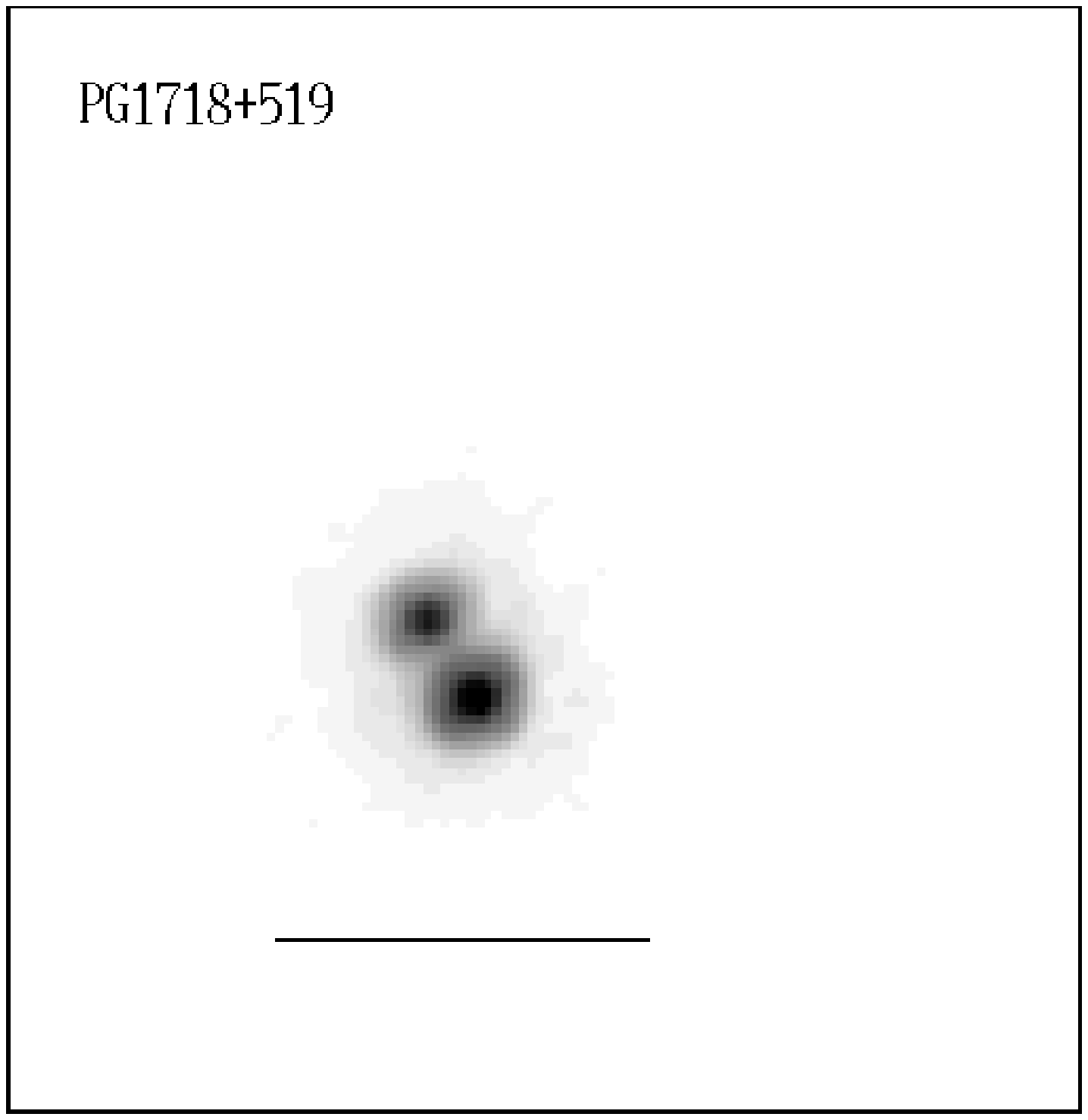}
\includegraphics{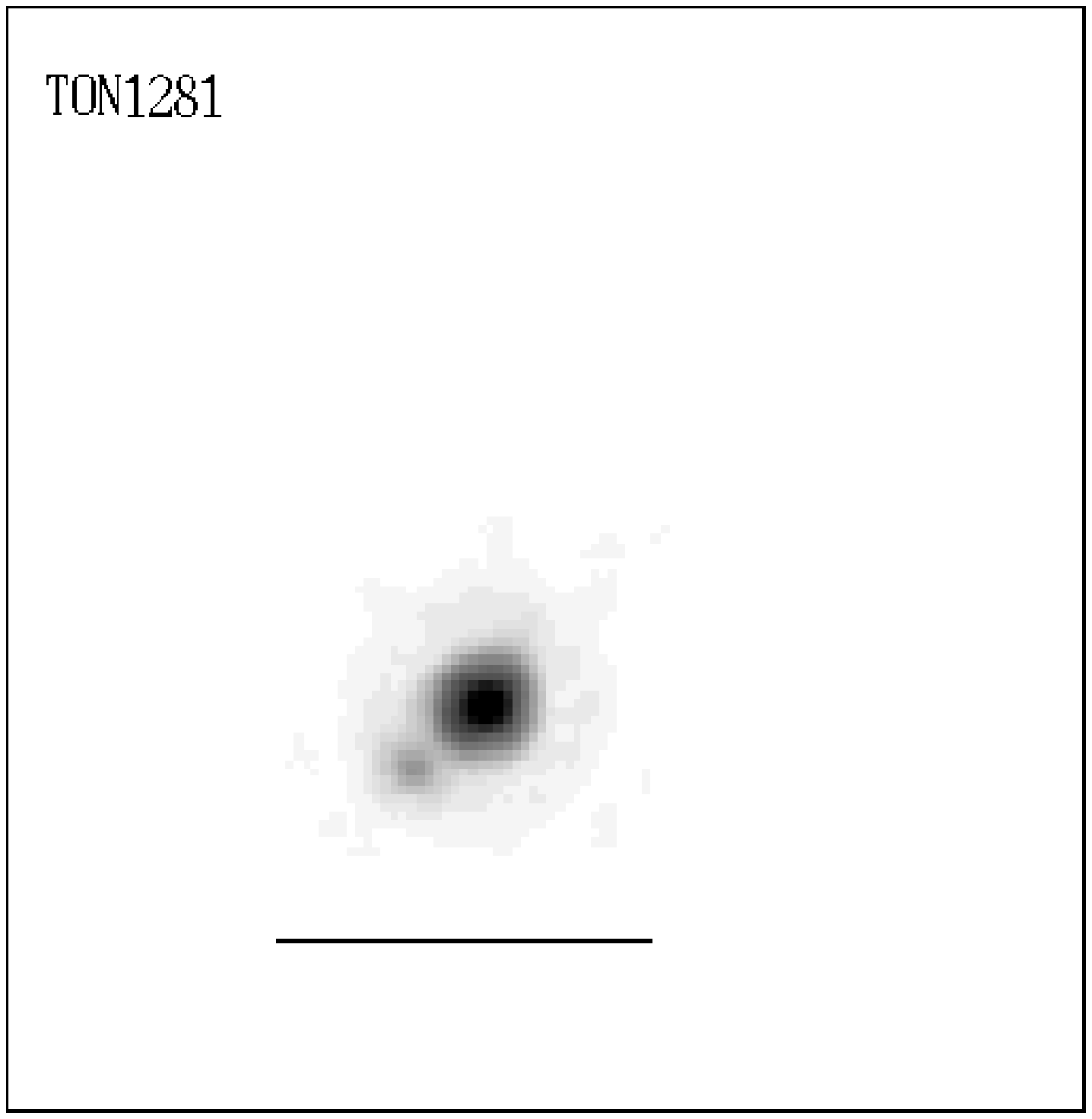}
\caption[]{The images of the resolved binaries. 
The bar in each image corresponds to 1\arcsec.
\label{resolv}}
\end{figure*}

The median-averaged images were first inspected by eye to see if any
companion could be detected. Only 6 stars (cf. Fig.~\ref{resolv}) showed
obvious nearby stars and the angular separations and brightness differences
can be found in Table~\ref{tab_bin}. The brightness differences were
determined using the command {\tt INTEGRATE/APERTURE} from 
{\tt MIDAS}, which performs
an aperture photometry with a given radius. Aperture photometry is 
difficult for TON~1281, TON~139, and PG~1718$+$519, due to the small
distance of the components. The sky background was determined in an empty
region using the same aperture as for the stars. 

\begin{table}
\caption[]{Separation and estimated brightness differences for the 
components of the 6 resolved binaries. The photometric data available for 
HE~0430$-$2457 do not allow to estimate a temperature or distance of the 
sdB. \label{tab_bin}}
\begin{tabular}{lrrl}
\hline
system & \multicolumn{2}{c}{separation} & brightness \\
       & angular & linear & difference\\
       &         & [AU] & $\Delta F675W$  \\
\hline
PG~0105$+$276 & 3\bsec37 & 3700 & \magpt{0}{9}\\
  & 4\bsec48 & 4900 & \magpt{1}{6} \\
HE~0430$-$2457 & 1\bsec25 & & \magpt{2}{1}\\
TON~1281 & 0\bsec22 & 250 & \magpt{3}{7}\\
TON~139 & 0\bsec32 & 300 & \magpt{0}{8}\\
PG~1558$-$007 & 2\bsec80 & 2500 & \magpt{3}{1} \\
PG~1718$+$519 & 0\bsec24 & 230 & \magpt{0}{8}\\
\hline
\end{tabular}
\end{table}

\begin{figure}
\vspace{7.5cm}
\includegraphics{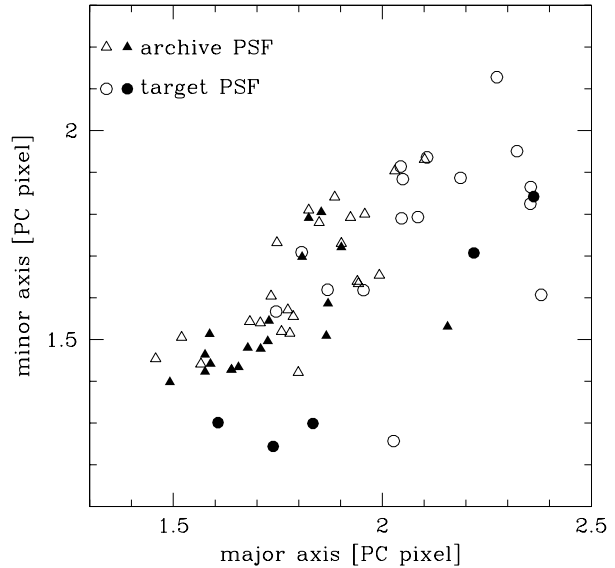}
\caption{The major and minor axes of the point spread functions for the 
target stars (circles, 
filled symbols mark brightest star of 
resolved binaries) and of archive point-spread 
functions (triangles, filled ones mark stars with positions on the PC 
chip close to our targets). 
\label{psf_axes}}
\end{figure}

\begin{figure*}
\vspace{11cm}
\includegraphics{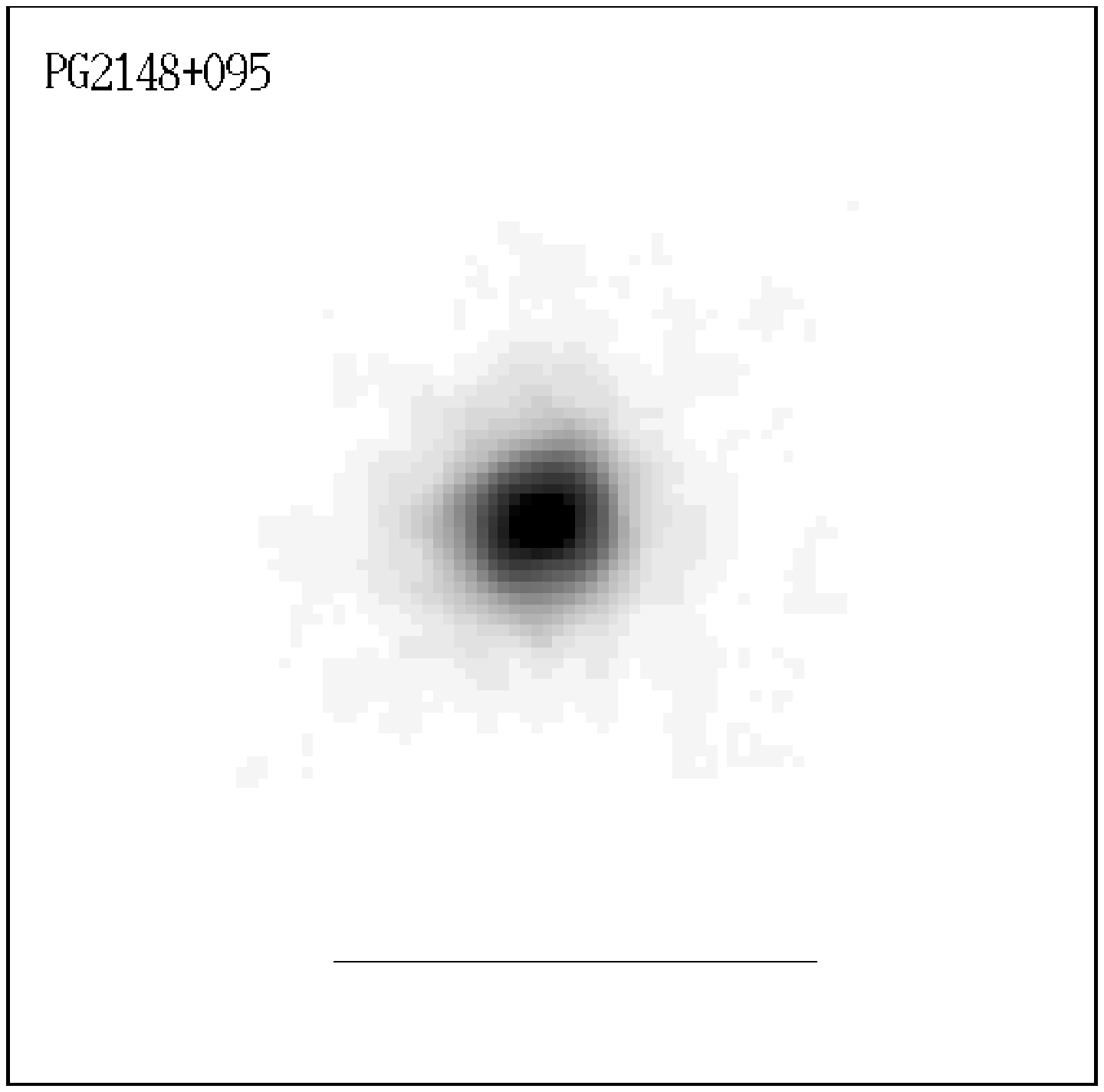}
\includegraphics{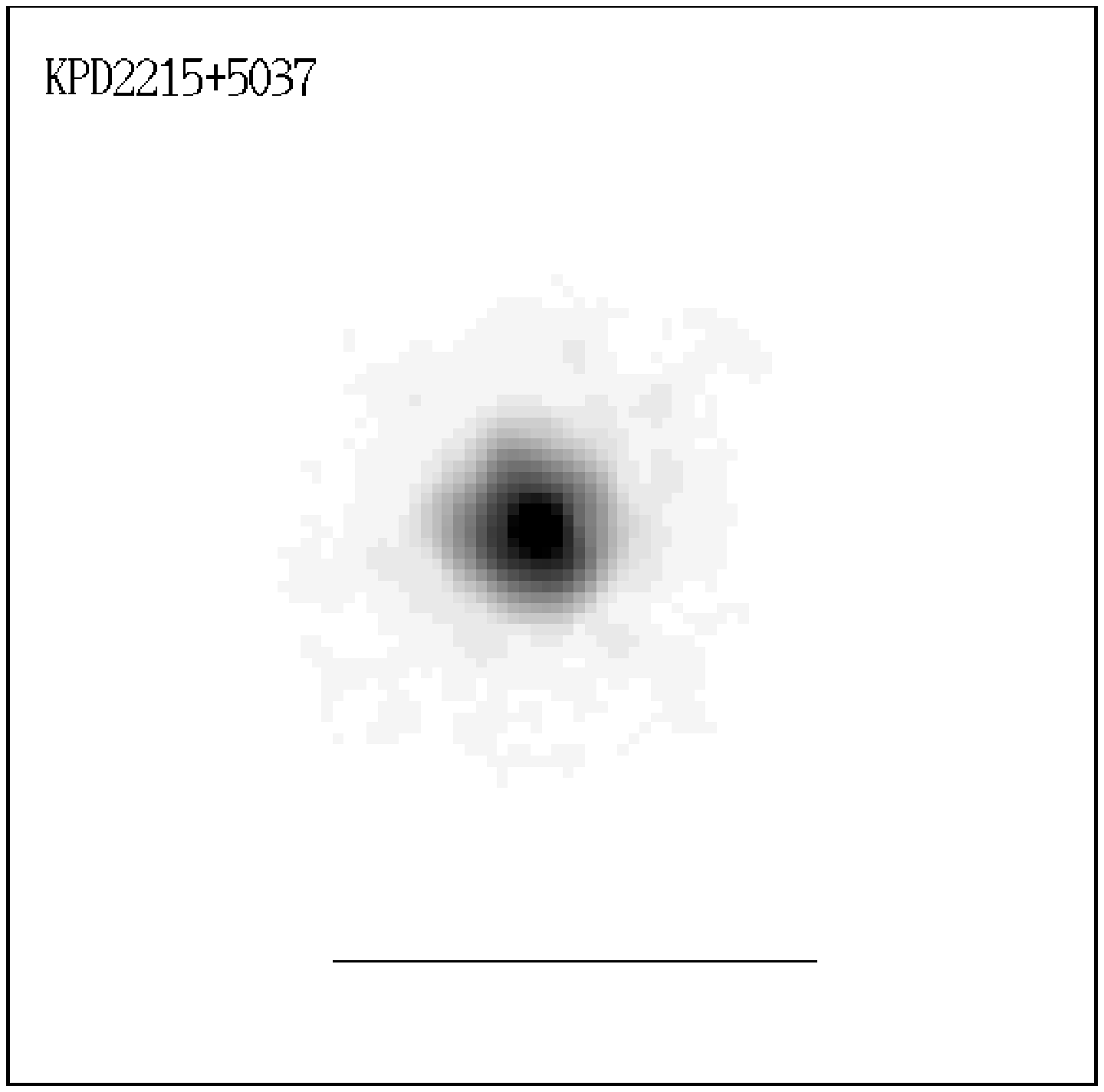}
\includegraphics{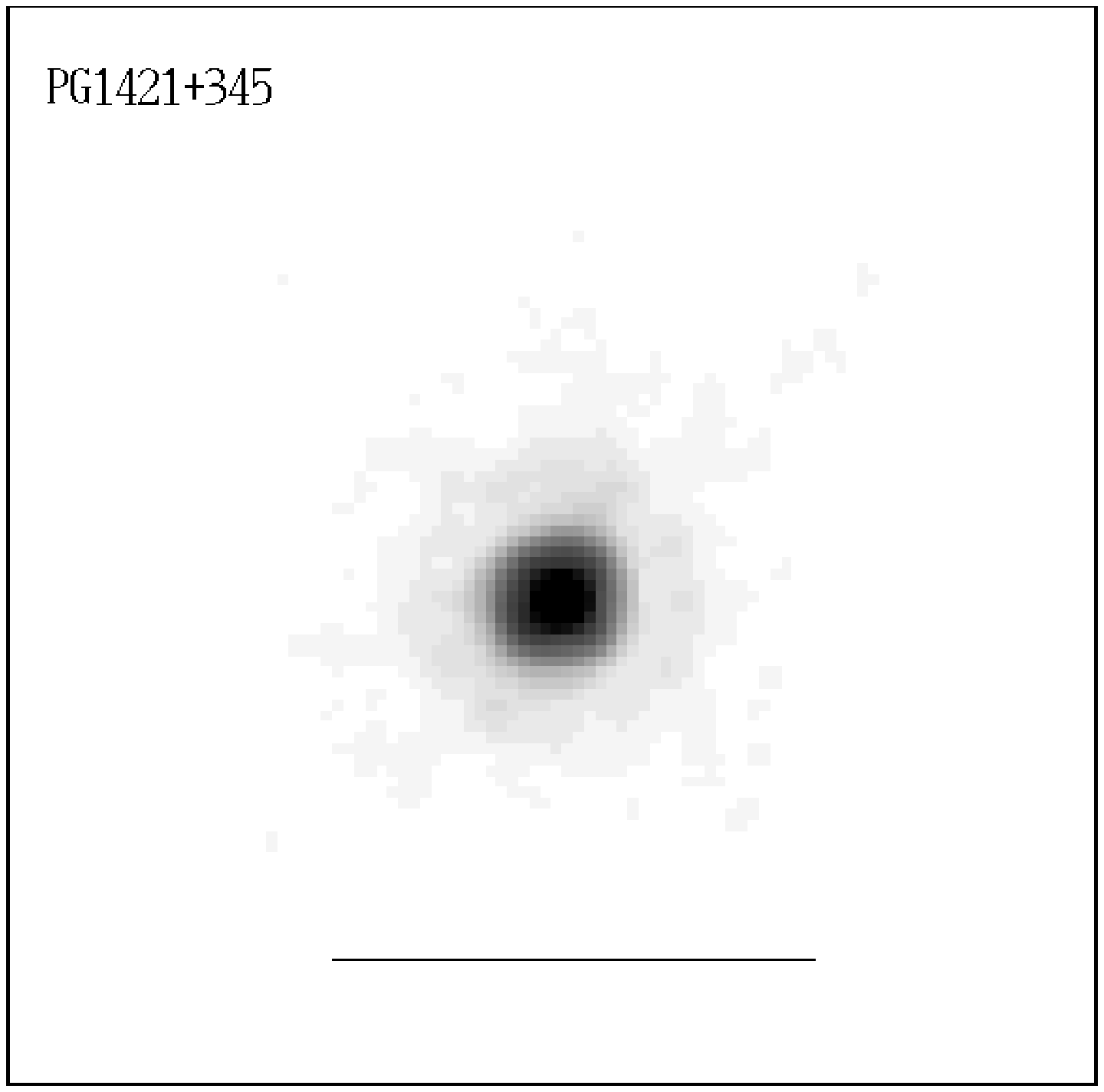}
\includegraphics{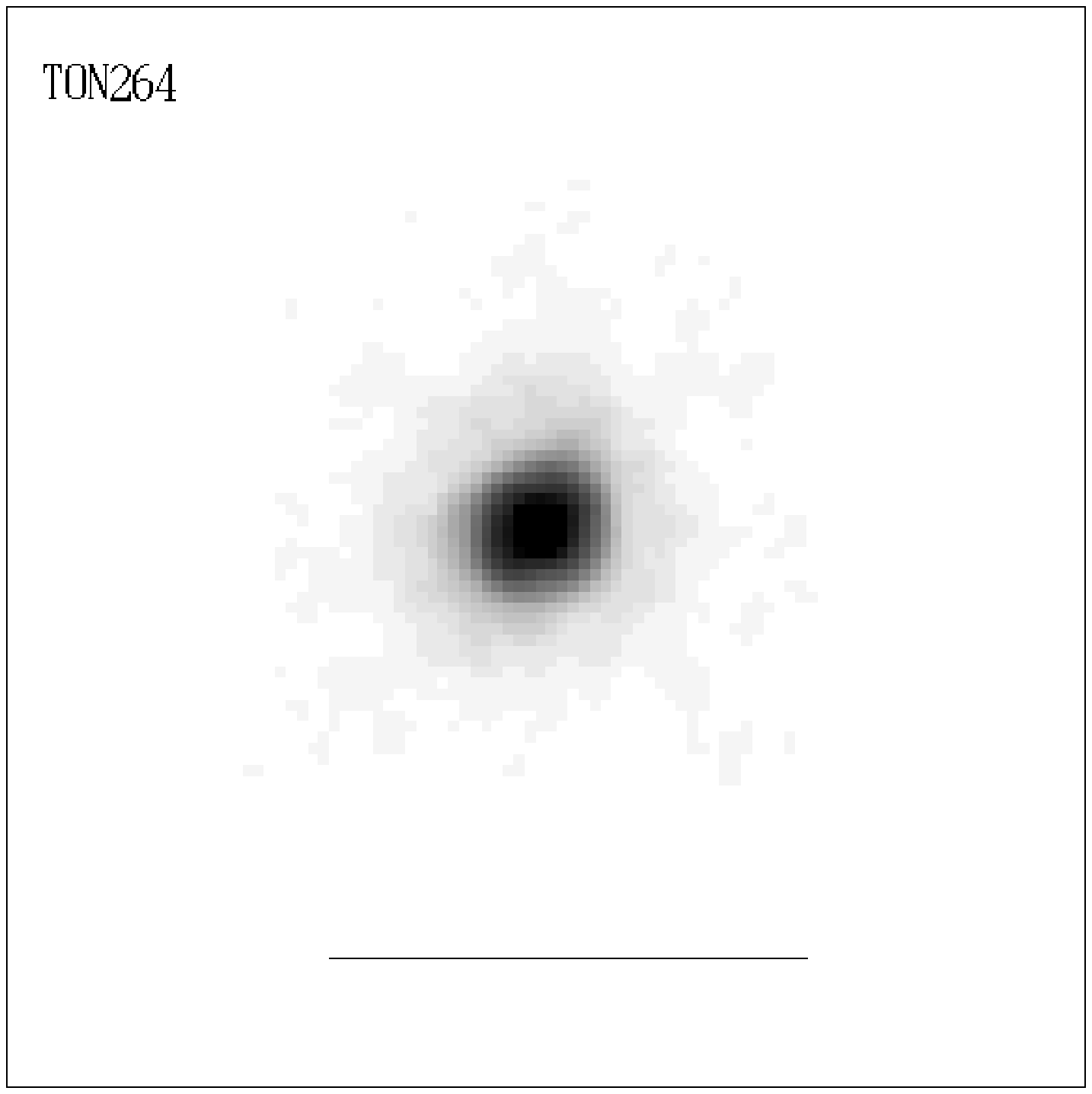}
\includegraphics{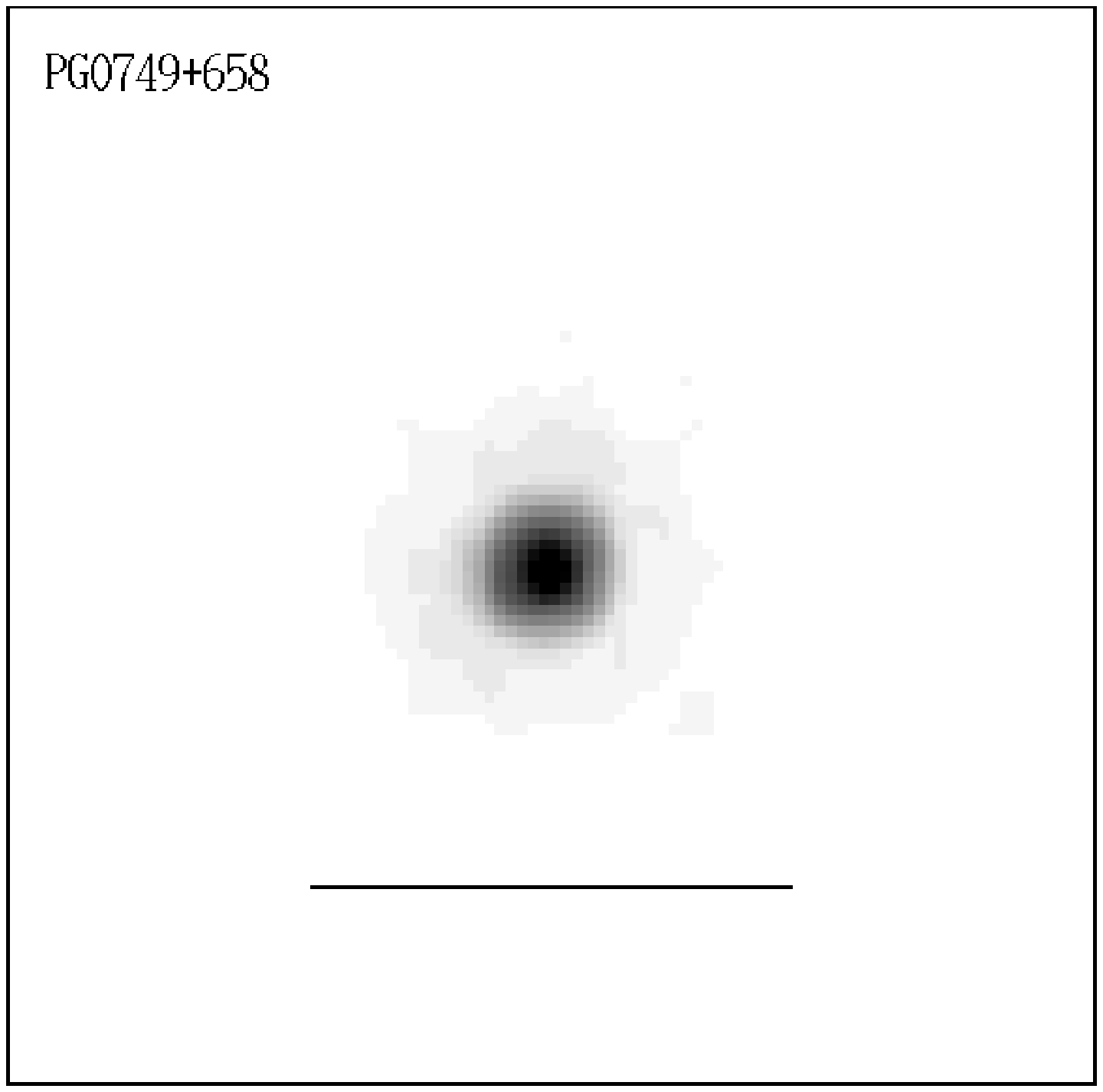}
\includegraphics{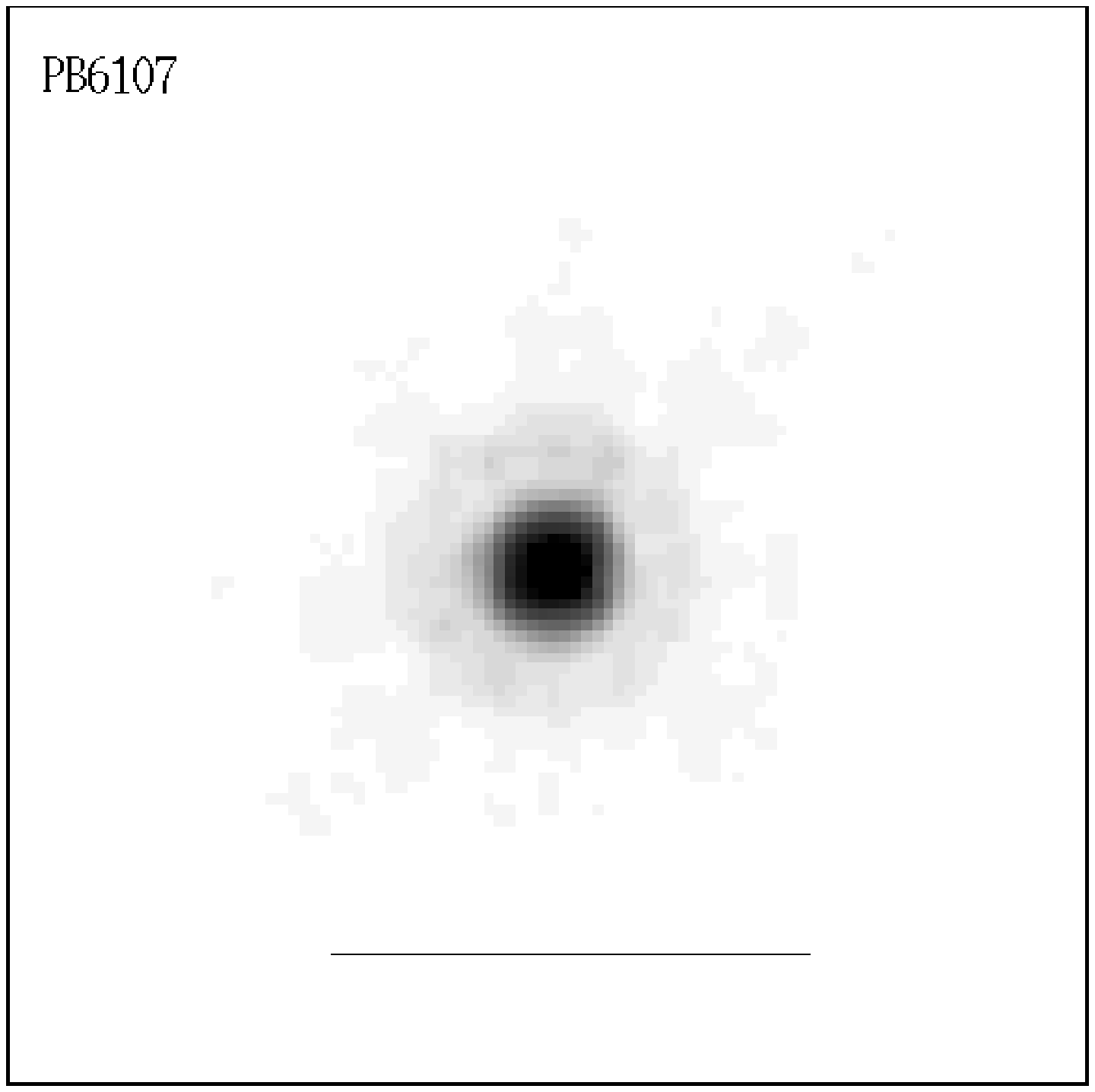}
\caption[]{The images of the unresolved stars (PG~2148$+$095,
KPD~2215$+$5037, TON~264, PG~0749$+$658) which show deviations from the 
standard PSF shape (see text). The images of PB~6107 and PG~1421$+$345 are well 
matched by the standard PSF shape and are displayed for comparison. Note that 
-- in contrast to all other stars displayed here --
there is no spectroscopic evidence for binarity of KPD~2215$+$5037.
The bar in each image corresponds to 1\arcsec.
\label{unresolv}}
\end{figure*}

To get a more quantitative estimate of possible companions we fitted
two-dimensional Gaussians with variable angle of the major axis to {\em
all} shifted and co-added target images and compared the results to fits
obtained for archive point-spread functions (PSFs; $F675W$ filter, PC
chip). The archive PSFs define a good correlation between the length of the
two axes, which is shared by most target PSFs (see Fig.~\ref{psf_axes}).
Besides the resolved binaries (where stray light can affect the
determination of the axis ratio) four stars deviate from the main
correlation between major and minor axis (see Fig.~\ref{unresolv}):
PG~2148$+$095 (2.03/1.26), KPD~2215$+$5037 (2.38/1.61), TON~264
(2.35/1.83), and PG~0749$+$658 (2.36/1.87). 

We used {\tt DAOPHOT} (Stetson \cite{daophot}) to obtain an average PSF
from those target stars that share the axis-relation of the archive PSFs.
This ``target PSF'' was then used to deconvolve all systems that are either
resolved by eye or show deviations from the axis-relation defined by
the archive PSFs. No additional components were resolved in this process,
but we could verify the brightness differences between the components of
the resolved systems listed in Table~\ref{tab_bin}, which were reproduced
by {\tt DAOPHOT} also for small separations. 

\begin{figure}
\vspace{7.5cm}
\includegraphics{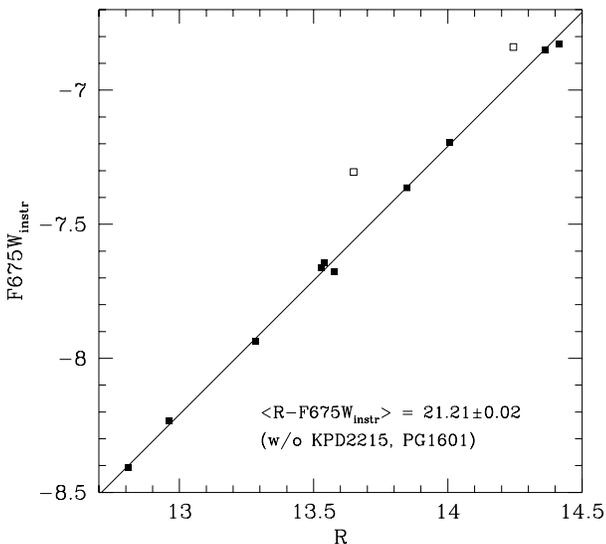}
\caption{The instrumental $F675W$ magnitudes compared to the $R_C$ data from 
Allard et al.\ (\cite{alwe94}). The open symbols are KPD~2215$+$5037 and 
PG~1601$+$145. The line marks the relation $R_C$ = 
$-2.5 \log 
{\frac{{\rm flux}_{0\bsec5}-{\rm sky}_{0\bsec5}}{\rm exposure \ time}}$+21.21.
\label{r_mag}}
\end{figure}

For 13 of our target stars a homogeneous set of ground-based $R_C$
measurements exist (Allard et al. \cite{alwe94}, see Table~\ref{broad}).
Comparing those data to the instrumental $F675W$ magnitudes integrated within
an aperture of 0\bsec5 radius 

\[ F675W = -2.5 \log 
{\frac{{\rm flux}_{0\bsec5}-{\rm sky}_{0\bsec5}}{\rm exposure \ time}} \]

\noindent
we find that most of the stars lie on a line with slope 1 (except 
KPD~2215$+$5037 and PG~1601$+$145, see Fig.~\ref{r_mag}). From the 11 stars 
on the line we determine a zeropoint of \magpt{21}{21} $\pm$ \magpt{0}{02}. 
From the WFPC2 data handbook we determine a zeropoint of 
\magpt{21}{9} (gain 14, including an aperture correction of \magpt{-0}{1})
that has to be corrected to Cousins $R$ by adding \magpt{-0}{65}
(assuming a spectral type of A5 for the combined spectra of our binary 
stars), yielding a final zeropoint of \magpt{21}{25}, in 
agreement with our empirically determined zeropoint. Since our empirically 
derived zeropoint automatically takes into account the unusual flux 
distribution of our binary stars we decided to use it to calculate $R_{HST}$
given in Table~\ref{broad}.

\section{Spectral energy distribution\label{sed}}

\begin{figure*}
\vspace{13.0cm}
\includegraphics{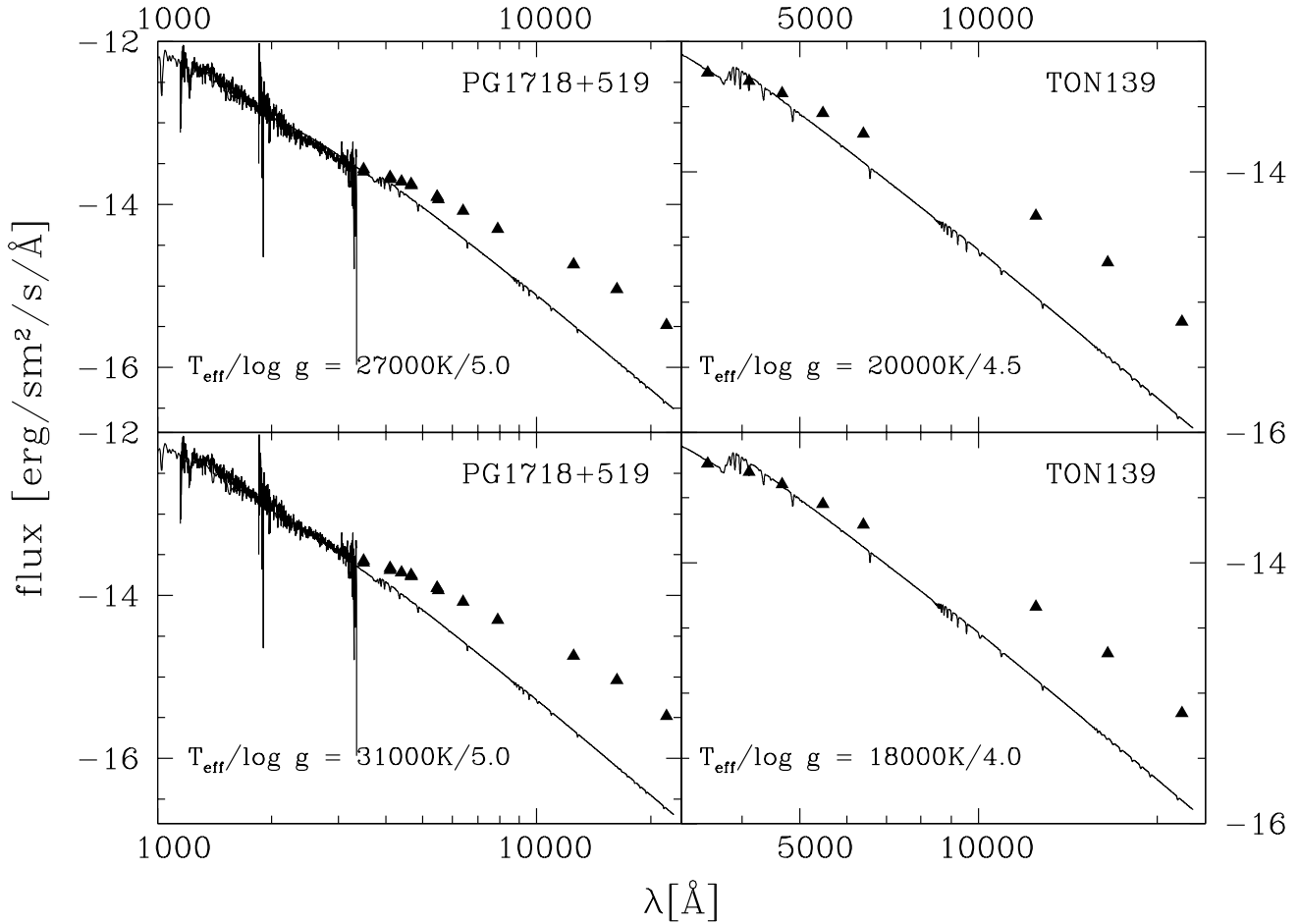}
\caption{Fits of ATLAS9 model spectra (Kurucz \cite{kuru92},
[M/H] = 0) to the photometric data of PG~1718$+$519 (left
panel, including IUE spectra) and TON~139 (right panel). The upper panels
show the fits obtained assuming that the bluest photometric data points
(IUE spectra and $u$ for PG~1718$+$519, $u$ and $v$ for TON~139) are not 
affected by the cool companion. The lower panels
show fits that reproduce the brightness differences measured on the WFPC2
images.
\label{fit_phot}}
\end{figure*}

To obtain an upper limit to our resolution we tried to estimate the $R$
brightness of the cool companion by fitting the available photometric data 
of those stars that have sufficient measurements.
In order to disentangle the flux of the hot star from
that of the cool star we analyse the composite spectral energy distribution.
For this purpose ultraviolet, optical and infrared (spectro-) photometry is 
collected from literature and archives (IUE, 2MASS). 
To determine the contribution of the hot star we fit synthetic 
spectra (Kurucz, \cite{kuru92}) to the bluest part of the observed spectral
range, i.e. 
IUE data plus $u$ or $u$/$U$ plus $v$/$B$ (if no UV data were available) and 
determine the effective temperature of the sdB star. In 
doing so we assume
that the companion does not contribute to the flux in this wavelength
range (cf. Fig.~\ref{fit_phot}). 
While this is probably true for the IUE data,  
some contamination may be present in 
the $u$/$U$- and $v$/$B$-band and consequently
the temperature determination for the sdB star can be compromised.

However, for some stars photometric data are so incomplete that no 
meaningful fit can be obtained.
Aside from the $F675W$ measurements discussed here PG~0942$+$461
and HE~2213$-$2212 have only $JHK$ photometry from 2MASS, which
are insufficient for a fit.
While HE~0430$-$2457 has $BVR$ photometry 
it is still not possible to constrain the sdB star's temperature with these 
data as $B-V$ is insensitive to \teff\ at sdB temperatures.
To convert the magnitudes into flux values we used the data given in 
Table~\ref{conversion}.

\begin{table}
\caption[]{Flux for a star with $m_\lambda$ = 0. The data are taken from 
Lamla (\cite{laml82}, p. 59, $uvby$; p. 82 $BVR_CI_C$), Zombeck 
(\cite{zomb90}, $JHK_{\rm UT98}$) and from the 2MASS Team
(priv.comm., $JHK_{\rm 2MASS}$).\label{conversion}}
\begin{tabular}{llr}
\hline
filter & flux & $\lambda_c$ \\
 & [erg/(cm$^2$ s \AA)] & [\AA] \\
\hline
$u$ & $1.169\cdot 10^{-8}$ & 3500 \\
$v$ & $8.444\cdot 10^{-9}$ & 4110 \\
$b$ & $5.826\cdot 10^{-9}$ & 4670 \\
$y$ & $3.700\cdot 10^{-9}$ & 5470 \\
$U$ & $4.187\cdot 10^{-9}$ & 3600 \\
$B$ & $6.597\cdot 10^{-9}$ & 4400 \\
$V$ & $3.607\cdot 10^{-9}$ & 5500 \\
$R_C$ & $2.254\cdot 10^{-9}$ & 6400 \\
$I_C$ & $1.196\cdot 10^{-9}$ & 7900 \\
$J_{\rm 2MASS}$ & $2.91\cdot 10^{-10}$ & 12510 \\ 
$H_{\rm 2MASS}$ & $1.11\cdot 10^{-10}$ & 16280 \\
$K_{\rm 2MASS}$ & $3.83\cdot 10^{-11}$ & 22030\\
$J_{\rm UT98}$ & $3.18\cdot 10^{-10}$ & 12500 \\
$H_{\rm UT98}$ & $1.18\cdot 10^{-11}$ & 16500 \\
$K_{\rm UT98}$ & $4.17\cdot 10^{-11}$ & 22000 \\
\hline
\end{tabular}
\end{table}

By comparing the measured flux in the R band to the model flux of the 
sdB star we derive the flux ratio of the hot vs. the cool star in the 
system.
For those systems 
which should have a rather bright companion according to their photometric 
data we verified the flux ratio in $R$ between sdB and cool companion 
from
two colour diagrams similar to those used by
Ferguson et al. (\cite{fegr84}), which is best 
suited for components of comparable brightness (for details see Ferguson et 
al. \cite{fegr84}). 
With this method 
we found that the companion of TON~1281 is bright enough to affect also the 
$u$ filter, yielding a temperature of 25,000\,K to 27,000\,K for the sdB 
instead of the 22,000\,K given in Table~\ref{tab-sep} and a brightness 
difference $\Delta R$ of \magpt{0}{2} to \magpt{-0}{1}. Also for PG~1601$+$345
we find a much smaller brightness difference (\magpt{0}{1}) and higher 
temperature (29,500~K) from this method than from our photometric fits. In 
this case the $B$ filter is already affected by the cool companion.
For reasons of consistency we keep the values from the photometric fits for 
these two stars in Table~\ref{tab-sep}.
For all other stars 
with brightness differences $\le$\magpt{0}{8} the results from both methods 
were the same. 
To correct for interstellar
reddening we used the reddening-to-infinity maps
of Schlegel et al. (\cite{scfi98}) which give
somewhat higher values than the older data of Burstein \& Heiles
(\cite{buhe82}). 
KPD~2215$+$5037, PG~1558$-$007, and
PG~2259$+$134 all lie in regions of quite high reddening according
to Schlegel et al. (\cite{scfi98}) and show no spectroscopic evidence
for a cool companion (see Appendix~\ref{app_spec}). The observed apparent 
infrared excess can be explained by high interstellar reddening alone, 
without invoking the presence of a cool companion. We also find no 
evidence for a companion from available photometry of PG~1656$+$213, 
although there is spectroscopic evidence (Ferguson et al, \cite{fegr84}).
However there are are no flux measurements 
redwards of V available and B and V fluxes 
are inconsistent. Therefore we keep PG~1656$+$213 as a programme star.

\begin{table*}
\caption[]{Estimated temperature of sdB stars, resulting reddening-free 
brightness of subdwarf B star ($R_{\rm sdB,0}$) and companion
($R_{\rm comp,0}$), distance d, brightness difference $\Delta R$, and
upper limit for linear separation a$_{\rm lim}$ 
derived from upper limit of angular 
separation $\Delta \alpha_{\rm lim}$. 
The reddening estimates are from the maps of Schlegel et al. (\cite{scfi98})
and we used $A_R = 2.6 \cdot$\ebv. 
The three different temperatures for PG~1511$+$624 result from
the three available SWP spectra. If no evidence for a 
companion can be found from available photometry no entry is given in column 
4.
\label{tab-sep}}
\begin{tabular}{l|rrrrrrrlr}
\hline
Star & $T_{\rm eff, sdB}$ & $A_R$ & $R_{\rm comp,0}$ & $R_{\rm sdB,0}$ & 
$M_{R, {\rm sdB}}$ & d & $\Delta R$ & $\Delta \alpha_{\rm lim}$ & 
a$_{\rm lim}$ \\
     & [K] &     & & & & [pc] & & & [AU]\\ 
\hline
PB~6107 & 23000 & \magpt{0}{086} & \magpt{14}{4} & \magpt{13}{0} & 
 \magpt{3}{3} & 870 & \magpt{1}{4} & 0\bsec1 & 87\\ 
PG~0105$+$276 & 32000 & \magpt{0}{156} & \magpt{15}{8} & \magpt{14}{4} 
 & \magpt{4}{4} & 1100 & \magpt{1}{4} & 0\bsec1 & 110 \\ 
PHL~1079 & 25000 & \magpt{0}{104} & \magpt{14}{9} & \magpt{13}{4} &
 \magpt{4}{4} & 630 & \magpt{1}{5} & 0\bsec1 & 63\\
PG~0749$+$658 & 22000 & \magpt{0}{125} & \magpt{14}{4} & \magpt{12}{1} 
 & \magpt{3}{3} & 580 & \magpt{2}{3} & 0\bsec2 & 116\\ 
%              & 25000 &  & \magpt{13}{6} &\magpt{12}{3} & \magpt{4}{4} 
% & 380 & \magpt{1}{3} & 0\bsec1 & 38\\ 
TON~1281 & 22000 & \magpt{0}{065} & \magpt{14}{4} &\magpt{13}{6} 
 & \magpt{3}{3} & 1150 & \magpt{0}{8} & 0\bsec07 & 80\\ 
TON~139 & 20000 & \magpt{0}{026} & \magpt{13}{6} & \magpt{13}{2} &
 \magpt{3}{3} & 950 & \magpt{0}{4} & 0\bsec05 & 48 \\
PG~1309$-$078 & 24000 & \magpt{0}{138} & \magpt{15}{5} & \magpt{14}{2}
 & \magpt{3}{3} & 910 & \magpt{1}{3} & 0\bsec1 & 91 \\
PG~1421$+$345 & 24000 & \magpt{0}{044} & \magpt{16}{0} & \magpt{14}{9}
 & \magpt{3}{3} & 2100 & 0\bsec1 & 210 \\
PG~1449$+$653 & 28000 & \magpt{0}{042} & \magpt{14}{7} &\magpt{14}{0} 
 & \magpt{4}{4} & 830 & \magpt{0}{7} & 0\bsec07 & 58\\ 
PG~1511$+$624 & 31000  & \magpt{0}{047} & \magpt{15}{7} &\magpt{14}{8} 
 & \magpt{4}{4} & 1200 & \magpt{0}{9} & 0\bsec07 & 84\\
              & 28000 &  & \magpt{15}{8} &\magpt{14}{8} & \magpt{4}{4} 
 & 1200 & \magpt{1}{0} & 0\bsec07 & 84\\
              & 33000 &  & \magpt{15}{6} &\magpt{14}{9} & \magpt{4}{4} 
 & 1260 & \magpt{0}{7} & 0\bsec07 & 88\\
PG~1558$-$007 & 23000 & \magpt{0}{468} & \multicolumn{1}{c}{} 
 & \magpt{13}{1} & \magpt{3}{3} & 910 & & & \\
PG~1601$+$145 & 25000 & \magpt{0}{133} & \magpt{15}{2} &\magpt{14}{6} 
 & \magpt{4}{4} & 1100 & \magpt{0}{6} & 0\bsec07 & 77\\
PG~1636$+$104 & 20000 & \magpt{0}{156} & \magpt{14}{5} &\magpt{13}{7} 
 & \magpt{3}{3} & 1200 & \magpt{0}{8} & 0\bsec07 & 84\\
PG~1656$+$213 & 17000 & \magpt{0}{172} & \multicolumn{1}{c}{}
 & \magpt{14}{6} & \magpt{3}{3} & 1800 & & & \\
TON~264 & 26000 & \magpt{0}{146} & \magpt{16}{0} &\magpt{14}{1} 
 & \magpt{4}{4} & 870 & \magpt{1}{9} & 0\bsec2 & 174\\
PG~1718$+$519 & 27000 & \magpt{0}{081} & \magpt{14}{1} &\magpt{14}{3} 
 & \magpt{4}{4} & 950 & \magpt{-0}{2} & 0\bsec05 & 48\\
% & 25000 & \magpt{0}{081} & \magpt{14}{4} &\magpt{14}{1} 
% & \magpt{4}{4} & 870 & \magpt{0}{3} & 0\bsec05 & 44\\
PG~2148$+$095 & 26000 & \magpt{0}{169} & \magpt{14}{5} & \magpt{13}{0} 
 & \magpt{4}{4} & 520 & \magpt{1}{5} & 0\bsec1 & 52\\
KPD~2215$+$5037 & 35000 & \magpt{0}{871} & \multicolumn{1}{c}{} 
 & \magpt{12}{8} & \magpt{4}{4} & 480 & & & \\
PG~2259$+$134 & 30000 & \magpt{0}{341} & \multicolumn{1}{c}{}
 & \magpt{14}{4} & \magpt{4}{4} & 1000 & & \\
BD~$-7^\circ$5977 & 29000 & \magpt{0}{093} & \magpt{10}{2}  & \magpt{11}{9}
 & \magpt{4}{4} & 320 & \magpt{-1}{7} & 0\bsec2 & 64\\
\hline
\end{tabular}
\end{table*}

Aznar Cuadrado \& Jeffery (\cite{azje01}) present an extensive discussion
of sdB parameters derived from energy distributions, which also includes
some of the stars discussed in this paper. In Table~\ref{compare_teff} we
present the temperatures given in their paper and other values collected
from literature in comparison to the ones derived here. As can be seen from
Table \ref{compare_teff} differences of $\pm$10\% in \teff\ between
different authors are quite common. 

\begin{table*}
\caption[]{Effective temperatures for sdB stars derived from energy 
distributions by various authors. The sources are Aznar Cuadrado \& 
Jeffery (\cite{azje01}, ACJ01), Allard et al. (\cite{alwe94}, A94),
Theissen et al. (\cite{thmo93},  T93;
\cite{thmo95}, T95), Ulla \& Thejll (\cite{ulth98}, UT98).
\label{compare_teff}}
\begin{tabular}{lllllll}
\hline
star & \multicolumn{6}{c}{\teff\ [K] derived by}\\
          & this paper & ACJ01  & T93    & A94    & T95    & UT98\\
\hline
PB~6107       & 23000  &       &       & 25000 &       & \\
PG~0105$+$276 & 32000  & 35850 &       & 32000 &       & \\
PHL~1079      & 25000  &       & 26350 &       & 30000 & 30000 \\
PG~0749$+$658 & 22000  & 25050 &       & 23500 &       & \\
TON~1281      & 22000  & 23275 &       & 29500 &       & \\
TON~139       & 20000  &       &       &       &       & 18000\\
PG~1449$+$653 & 28000  & 28150 &       & 28000 &       & \\
PG~1511$+$624 & 31000: &       &       & 33000 &       & \\
PG~1636$+$104 & 20000  &       &       & 21000 &       & \\
TON~264       & 26000  &       &       & 28500 &       & \\
PG~1718$+$519 & 27000  & 29950 & 23500 & 25000 & 30000 & \\
PG~2148$+$095 & 26000  & 22950 &       & 26000 &       & 25000 \\
KPD~2215$+$5037 & 35000 &      &       & 24500 &       & \\
PG~2259$+$134 & 30000  & 28300 & 28500 &       & 22500 & \\
\hline
\end{tabular}
\end{table*}

The temperatures derived from the photometric data and from line profile
fits for the stars in regions with high reddening agree moderately well
(compare Tables~\ref{tab-sep} and \ref{tab_spec}). The discrepancies may be
due to small scale variations in reddening that affect the temperatures
derived from photometry but not those derived from line profile fits. 

From the photometric fit we can derive the apparent 
$R$ magnitudes of the sdB and of
the cool star and correct both for interstellar extinction. The
uncertainty in \teff\ of about $\pm$10\% evident from
Table~\ref{compare_teff} causes an estimated uncertainty in the derived
brightness for both components of \magpt{\pm0}{2}. Knowing the absolute $R$
magnitude of the sdB stars then allows to determine their distance. We use
the mean $M_V$ derived by Moehler et al. (\cite{mohe97}) for hot subdwarfs
in the globular cluster NGC~6752. They found two groups of hot subdwarfs, a
cooler one with a mean effective temperature of 22,000\,K and $<M_V>$ =
\magpt{3}{2} (5 stars), and a hotter one with $<$\teff$>$ = 29,000\,K and
$<M_V>$ = \magpt{4}{2} (12 stars). From Kurucz (\cite{kuru92}) model
atmospheres for [M/H] = 0 we find $V-R$ = \magpt{-0}{120} for \teff\ =
22,000\,K and \magpt{-0}{152} for 29,000\,K. We therefore use $M_R$ =
\magpt{3}{3} for stars cooler than 25,000\,K and $M_R$ = \magpt{4}{4} for
hotter stars. 

Using the archive point spread functions we estimated the minimum separation
that we can resolve for a given brightness difference by adding two PSFs
with a defined brightness difference and angular separation and examining
the resulting image by eye. We find the following resolution limits:
$\Delta \alpha_{\rm lim}$ ($\Delta R$) = 0\bsec2 (\magpt{2}{0}), 0\bsec1
(\magpt{1}{5}), 0\bsec07 (\magpt{1}{0}), 0\bsec05 (\magpt{0}{5}). Using the
distances determined above we can now derive upper limits for the linear
separation of the unresolved binaries (cf Table~\ref{tab-sep}), ranging 
from 50~AU to 210~AU. 

%\section{Discussion\label{disc}}

Table~\ref{tab_bin} shows that the brightness differences between the components
in  TON~1281 and HE~0430$-$2457 are too large to reproduce the spectral 
energy distribution of TON~1281 and the photometry of HE~0430$-$2457, 
respectively.
The large 
brightness difference of
\magpt{3}{1} (from the WFPC2 data) for PG 1558$-$007 agrees with the lack
of photometric and spectroscopic
evidence for a companion. 
In the remaining two cases
(PG~1718$+$519, TON~139) 
%the resolved components are bright enough to produce a detectable infrared
%excess, but
%For PG~1718$+$519 and TON~139 the observed brightness differences
the brightness differences in Table~\ref{tab_bin} are somewhat larger than 
those derived from the spectral energy distribution. 
To see 
whether we can in principle accommodate the HST observations by fits to the 
photometric data we repeated the fits, this time enforcing the brightness 
difference in the R band obtained from the HST data. The results are 
shown in 
Fig.~\ref{fit_phot} (in comparison to the original fits). Obviously the 
companion of PG~1718$+$519 is sufficiently bright to affect also the $u$ 
filter, thereby rendering our assumption that this filter is unaffected by 
the cool companion obsolete. The fits for TON~139 do not show much 
difference. We conclude that the spectral energy distribution of TON~139 
and PG~1718$+$519 are consistent with the R band flux ratio measured with 
the HST WFPC2 camera.   

\subsection{The sdO star PG~0105$+$276}

Since the He-sdO PG~0105$+$276 does not 
belong to the programme sample, we discuss it 
separately. It is the only programme star that is resolved into three 
components. However, the two companions are quite distant from the primary
(3\bsec37 and 4\bsec48, respectively).
The light of these companions can explain at least qualitatively 
the IR excess observed by ground based aperture photometry. The spectrum of 
PG~0105$+$276, however, does not show any signature of a cool companion, 
probably because 
due to the orientation and the small width of the slit no light of the 
distant companions was included. 
The diaphragm used in the photometry was large
(18\arcsec) and included the companions' light.   

%In addition the helium abundance is much larger than the solar one adopted for 
%the photometric fits.

The brightness differences measured on the WFPC2 image (\magpt{0}{9}, 
\magpt{1}{6}) for PG~0105$+$276
are smaller than the one derived from the photometric fit
(\magpt{1}{4}), i.e. one companion is brighter than expected. However,
as discussed in Appendix~\ref{app_spec}, 
the true temperature (from line profile fitting) is much 
higher than the one obtained from the spectral energy distribution 
(63,000\,K vs. 35,000\,K)
making the companion's luminosity obtained from photometry a lower limit 
only.

%Restricting ourselves to those sdB stars that do show 
%evidence for a cool companion
%(by a photometric red excess and/or in their spectra) we
%exclude PG~0105$+$276, PG~1558$-$007, KPD~2215$+$5037, and 
%PG~2259$+$134 from any statistical discussion 
%and are left with a sample of 19 stars.

\section{Simulation of separability in binary systems}

In order to interpret our results with respect to the different 
evolutionary scenarios we simulate binary systems containing
main sequence (MS) companions and sdBs with period distributions found for
normal main sequence binaries (Duquennoy \& Mayor \cite{duma91}).  Assuming
that the sdB mass is 0.5\Msolar\ and the MS companion mass is 1\Msolar\ we
convert the period distribution published by Duquennoy \& Mayor (\cite{duma91})
to physical separations using Kepler's
Harmonic law. The orientation of the axis of the system is then chosen to
be random in space and the projected separation, or $a \sin i$, is
calculated, given the distance to the system which is found from the
apparent and absolute brightness of the system. The orbits are assumed to 
be circular. 

%\begin{figure}
%\vspace{6.5cm}
%\special{psfile=thejll_fig.ps hscale=55 vscale=55 hoffset=-50 voffset=-200
%angle=0}
%\caption{The predicted cumulative distribution of binary separations, using 
%the data from %Heintz \cite{hein69}, dashed line) and 
%Duquennoy \& Mayor (\cite{duma91}, solid line).
%\label{thejll}}
%\end{figure} 

%For simplicity we assume that the visual magnitude of the system is
%given by the sdB alone. This is somewhat justified by the experience
%from visual and $JHK$ IR observations of hot subdwarf systems.  Some, but
%not all, systems are characterized by visual magnitudes set by the sdB
%while the IR magnitudes can be dominated by companions.
Based on the spectroscopic distances derived above (see Table~\ref{tab-sep})
we then simulate a huge number of such binary systems. For three stars 
(HE~0430$-$2457, PG~0942$+$461, and HE~2213$-$2212) the
magnitude ratio of the components could not be determined and therefore the 
distances are unknown. We adopted the mean value of the other stars 
($\Delta R$ = \magpt{1}{1}), 
which is consistent with their spectral appearance (see Fig.~\ref{bin_spec}).
The numerical simulation predicts a mean value of $a \sin i$ = 0\bsec04 and
that, out of the 19 observed systems, 
we should resolve six systems at a resolution limit of 0\bsec1, 
one of which should show a separation greater than 1\bsec0.

Since the orbital motion for an eccentric orbit is lower during phases 
of large separation, the time averaged distance is larger than the semi 
major axis. Thus eccentric orbits would increase the detectability.
Duquennoy \& 
Mayor (\cite{duma91}) also provide a distribution of ellipticities for 
normal stars. If the sdB systems did not experience phases of binary 
interaction, the distribution of eccentricity should correspond to that of 
normal stars. We used Duquennoy \& Mayor's distribution corrected for 
selection effects. For each eccentricity the ratio of the time averaged 
distance to $a$ was calculated and finally the mean over the Duquennoy \& 
Mayor distribution computed.  
We find the average 
distance of the companions to increase by 17\%.
Another mechanism that tends to increase the separation of the components
in a sdB binary is mass loss during post-main sequence evolution
in order to reduce the mass of the 
sdB progenitor to its present value of half a solar mass. Assuming that
the sdB evolved from a 1\Msolar\ main sequence progenitor it must have lost 
0.5\Msolar\ due to a stellar wind during its post-main sequence evolution. 
Assuming that the wind emanates in a spherical symmetric manner and does 
not interact with the companion the increase in separation can be 
calculated according to $\frac{\dot{a}}{a}=-\frac{\dot{M_s}}{M_s+M_c}$ 
(Pringle \cite{prin85}), with $a$ being the separation and $M_s$ and 
$M_c$ the masses of the 
sdB progenitor and that of the cool star, respectively. As a result
the separation increases by 33\%. 

We repeated the Monte Carlo simulations for increased separations. Even 
when we consider both elliptical orbits and evolution of the orbits due 
to a stellar wind as described above the prediction increased only slightly 
7 resolvable stars in our sample.

Hence we predict that 6 to 7 stars should be resolvable in our sample if 
the systems have separations consistent with the Duquennoy \& 
Mayor (\cite{duma91}) distribution.

\section{Chance projections and triple systems}

In the vicinity of five programme stars we found an additional object within
a radius of 3\bsec0\footnote{Note that PG~0105+276, which is resolved in three 
components 
(see Fig.~\ref{resolv}), is an sdO star and does not belong to our sdB sample.}.
We have demonstrated above that only in two cases (TON~139 and 
PG~1718+519) the relative brightnesses are 
consistent with the expectations
from the deconvolution of the spectral energy distribution. The remaining
three cases must then be chance projections or triple systems.
Since the programme stars lie at high galactic latitudes (except 
KPD~2215$+$5037, see Table~\ref{tab_targ}), 
we expect chance coincidences to be rare. Indeed, we do not find any
additional object in the PC field (40\arcsec$\times$40\arcsec) except for
the low galactic latitude object KPD~2215$+$5037. 

According to Abt \& Levy (\cite{able76}) 16\% of multiple systems of normal 
stars are 
triples. If the fraction of triple systems is the same for our sample,
we expect three programme stars to be triple.
Most of these, if not all, should be resolvable. Besides TON~139 and 
PG~1718$+$519 we find in three cases companions to the sdB stars which are 
too faint to match the spectral energy distribution. These could be triple 
systems consisting of an unresolved sdB binary and a distant third star. 

\section{Radial velocities}

\begin{table*}
\caption[]{Heliocentric radial velocities for the sdB- and the cool star 
components of TON~139 and PG~1718+519.\label{radvel}}
\begin{tabular}{llrrrrr}
\hline
star & date & HJD-2450000 & exposure & S/N &  $v_{\rm rad}$ 
[km\,s$^{-1}$] & 
$v_{\rm rad}$ [km\,s$^{-1}$] \\
& UT & & time [s] & & (sdB component) & (cool companion)\\
\hline
 TON~139 & 1996-01-14 &  96.91396 & 600 & 95.6 & $-$6.3$\pm$4.9 & 19.9$\pm$0.6\\
 TON~139 & 1996-03-11 & 153.84515 & 300 & 71.9 & $-$7.4$\pm$7.8 & 20.2$\pm$0.7\\
 TON~139 & 1996-06-09 & 243.75586 & 600 & 66.5 & $-$13.1$\pm$8.4 & 21.8$\pm$0.8\\
 TON~139 & 1997-01-28 & 476.96939 &1800 & 69.4 & $-$20.2$\pm$7.9 & 20.2$\pm$0.9\\
 TON~139 & 1997-07-04 & 633.66734 & 500 & 72.7 &  32.6$\pm$6.7 & 22.4$\pm$0.7\\
 TON~139 & 1998-01-22 & 836.03834 & 750 & 82.9 & $-$22.1$\pm$9.1 & 20.7$\pm$0.6\\
 TON~139 & mean       &           &     &      & $-$3.6$\pm$20.2& 20.8$\pm$1.0\\
 PG1718+519 & 1997-09-10 & 701.71120 &1400.0 & 82.0 &$-$69.2$\pm$10.1& $-$68.0$\pm$0.9\\
\hline                                               
\end{tabular}
\end{table*}

Important additional information can be obtained from radial velocity 
measurements. A systematic search for
radial velocity variations of our programme stars is 
needed. Such projects have already been started by Saffer et al.
(\cite{sagr01}) and Maxted et al. (\cite{mahe01}) 
who observed six of our
programme stars (PB~6107, PHL~1079, PG~0749$+$658, TON~1281,
PG~1449$+$653 and PG~2148$+$095). None of them showed 
significant radial velocity changes.

Saffer et al. (\cite{sagr01}) find in their survey of 21 composite
spectrum sdB stars that the velocity variations of the individual
components as well as the velocity difference between the two components
are very small (less than a few km\,s$^{-1}$) or undetectable, and conclude that
the binaries have likely periods of many months to several years. Green et 
al. (2001) estimate from 
these measurements that the current periods average 3 -- 4 years with 
separations 540 -- 650 \Rsolar. 

We have obtained multiple precise radial velocities for TON~139 and a 
single measurement of PG~1718$+$519 using the MMT Blue Channel spectrograph 
at 1\AA\ resolution from 4000-4930\AA\ (see table~\ref{radvel}). 
The radial velocities of the cool companions were determined by 
cross correlation against super-templates of main sequence spectral types 
from F6 to K5.
The sdB velocities were derived using a preliminary attempt at subtracting 
out the cool star companion spectrum. For details 
of the data reduction and analysis see Saffer, Green \& 
Bowers (\cite{sagr01}). 
Improved sdB velocities using better cool star template spectra for 
the subtractions will be determined by Green, Bowers \& Saffer 
(2002, in prep.).

For TON~139 the cool star's velocity is 
constant, whereas the sdB velocity is changing by more than 
50\,km\,s$^{-1}$.
This can be explained if an additional companion is orbiting 
the sdB star. This 
companion has to be so faint that it does not contribute to the light in 
the R band. Hence we have to conclude that the resolved system TON~139 
is a triple system.
A radial velocity study of PG~1718$+$519, the second resolved system in our 
sample, is not available yet. The single 
measurement listed in Table~\ref{radvel} gives identical radial velocities 
for the sdB and the cool companion. 
This argues against a third faint component 
orbiting the sdB star in a narrow orbit as was found for TON~139. Additional 
radial velocity measurements are urgently needed to clarify the nature of 
PG~1718$+$519. Assuming that PG~1718$+$519 is not triple, this would be the 
only resolved 
binary system in our sample of 19 objects.

\section{Conclusions}

In total we have resolved six systems out of a sample of 23 stars. Of those 
23 stars, however, four do not really belong to the intended sample of sdB 
stars showing evidence for a cool companion: PG~1558$-$007, KPD~2215$+$5037,
and PG~2259$+$134 show no photometric or spectroscopic evidence for a 
companion. The observed infrared excess can be explained by 
interstellar reddening rather than by a cool companion.

PG~1558$-$007 does have a resolved near by star (linear separation 1500\,AU), 
which, however, is too faint to contribute detectably to the combined light
in the R band. 
PG~0105$+$276 is a helium-rich sdO star (with two possible distant
companions at 3700\,AU and 4900\,AU). 

Of the remaining four resolved systems the nearby stars
are in two case (TON~1281, HE~0430$-$2457) too faint to reproduce the 
photometric and/or spectroscopic observations of the stars.

Only in the two systems TON~139 and PG~1718$+$519 
(separations 0\bsec32 and 0\bsec24, respectively) 
do the magnitudes of the resolved components match the expectations.
These two stars could be physical binaries whereas in the other cases 
the nearby star may be a third component or a chance projection.
Radial velocity measurements indicate, however, that the resolved system
TON~139 is also triple.

Hence, the observed sdB binary sample was reduced to 19 objects with 
two bona-fide resolved systems, which have apparent separations of 
0\bsec24 and 0\bsec32.
From the numerical simulations 
we would expect to resolve
six to seven systems if sdB stars have the same binary 
characteristics as normal stars, out of which one system is expected to 
have $a \sin i>$1\arcsec\ and two should have separations between 0\bsec1 and 
0\bsec2.
The discrepancy becomes even more pronounced if one recalls that our
photometric fit procedure tends to underestimate the brightness of the
companion (and thus to overestimate the limiting angular separation that
can still be resolved). 
In addition we expect three triple systems to 
be present in our sample. Most of these, if not all, should be resolvable.   
Such systems could explain some of 
the more distant companions as well as the 
radial velocity measurements of TON~139.

This success rate (1 resolved binary out of 19 candidates) is clearly {\it
below} the prediction of numerical simulations assuming single star
evolution (about 30\%), using the distribution of binary separations 
given by Duquennoy \& Mayor (\cite{duma91}). This indicates that
the distribution of separations of sdB binaries strongly deviates from that
of normal stars. 

If, on the other hand, all sdB stars were produced by close binary evolution,
none of the binary systems should have been resolved (even at the high
spatial resolution of the WFPC2 camera).
Our low success rate is thus closer to that predicted by the close binary 
evolutionary scenario. 
Recent radial velocity surveys (Saffer et al.
\cite{sagr01}; Maxted et al. \cite{mahe01}) revealed that a large fraction
of single-lined sdB stars are indeed close binaries with periods below 10 days.
Our results could be explained if most of the programme stars were 
close binaries.
Therefore, our study provides further evidence that close binary evolution 
indeed is fundamental to the evolution of sdB stars. A survey for radial 
velocity variations in all of our programme stars will be tale telling. 

\begin{acknowledgements} 
This work was supported by the DLR under grant 50\,OR\,96029-ZA. We thank
Klaas de Boer (Bonn), Heinz Edelmann (Bamberg) and Heinz Lehnhart 
(T\"ubingen) for taking most of the optical spectra for us 
and Martin Altmann (Bonn) for providing us with his photometric 
measurements prior to publication. Thanks go also to Anna Ulla and Klaas 
de Boer for helpful comments and encouragement. 
This publication makes use of data
products from the Two Micron All Sky Survey, which is a joint project of
the University of Massachusetts and the Infrared Processing and Analysis
Center/California Institute of Technology, funded by the National
Aeronautics and Space Administration and the National Science Foundation.
We also made extensive use of the Simbad database, operated at CDS,
Strasbourg, France. 
\end{acknowledgements}

\clearpage

\appendix
\section{Spectroscopic observations and data reduction
 \label{app_spec}}

\begin{table*}[t]
\caption[]
{New optical spectroscopy and atmospheric parameters of single programme 
stars\label{tab_spec}}
\begin{tabular}{lrrrrrrr}
\hline
star & telescope and & wavelength & spectral & obs. date & \teff\ & \logg\ & 
 log(He/H) \\
 & spectrograph & range  & resolution  & & & \\     
 & & [\AA] & [\AA] & & [K] & [cgs] & \\     
\hline
PG~0105$+$276& CA 3.5m TWIN & 3600 -- 7400 & 3.1 & 1997/08/31 & 
63000 & 5.4 & $+$0.5 \\ 
HE~0430$-$2457 & ESO 1.5m B\&C & 3600 -- 7450 & 5.5 & 1996/10/22 & & & \\
PG~0942$+$461 & CA 3.5m B\&C  & 3860 -- 5560 & 5.0 & 1989/01/23 & & & \\
PG~1309$-$078  & ESO 1.5m DFOSC & 3860 -- 6780  & 5.4 & 2000/06/21 & & & \\
PG~1558$-$007  & ESO 1.5m DFOSC & 3860 -- 6780  & 5.4 & 2000/06/21
& 20300 & 5.0 & $-$2.6 \\
PG~2148$+$095  & ESO 1.5m B\&C & 3730 -- 4970 & 3.0 & 1991/07/10-15 & & & \\
HE~2213$-$2212 & ESO 1.5m B\&C & 3600 -- 7400 & 5.5 & 1996/10/23  & & & \\
KPD~2215$+$5037 & CA 3.5m TWIN & 3260 -- 7450 & 3.1 & 1997/08/29 
& 29400 & 5.6 & $-$2.2\\
PG~2259$+$134 & \multicolumn{4}{c}{Theissen et al. \cite{thmo93}} 
& 31900 & 5.9 & $-$1.7\\
\hline
\end{tabular}
\end{table*}

The observational setups and observing dates for the new spectra are given 
in Table~\ref{tab_spec}. The reduction of the spectra of PG~0105$+$276,
HE~0430$-$2457, PG~0942$+$461, HE~2213$-$2212, and KPD~2215$+$5037 
are described by 
Edelmann et al. (\cite{edel01a}). PG~2148$+$095 was observed and reduced 
as described by de Boer et al. (\cite{dbsc95}), the reduction of 
PG~1309$-$078 and PG~1558$-$007 was performed in the same way as described in 
Moehler et al. (\cite{mohe97}). 

\begin{figure}
\vspace{7.5cm}
\includegraphics{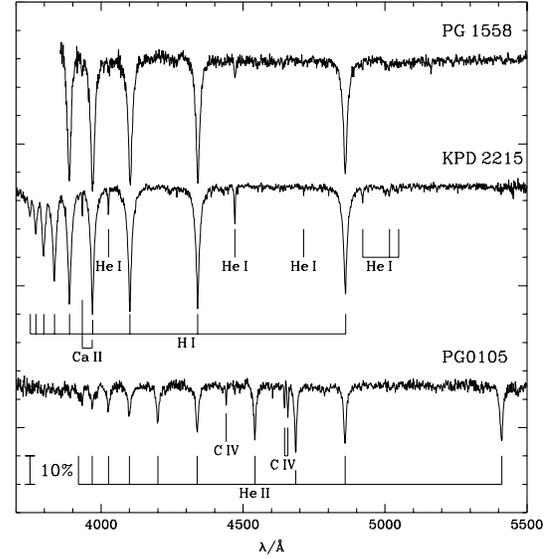}
\caption{Optical spectra of the sdB stars PG~1558$-$007 and 
KPD~2215$+$5037 as well as of the sdO star PG~0105$+$276. The spectra of 
the former are dominated by hydrogen lines whereas that of the latter 
by \ion{He}{ii} lines.\label{sing_spec}}
\end{figure}

Fig.~\ref{sing_spec} shows the spectra of the stars that show no
spectroscopic or photometric evidence for a cool companion
(PG~1558$-$087, KPD~2215$+$5037, and PG~0105$+$276).
The \ion{Ca}{ii} absorption lines in the 
spectra of these stars  (see Fig.~\ref{sing_spec}) 
are probably of interstellar nature.
Our spectrum clearly shows that PG~0105$+$276 is a 
helium rich sdO star (see Fig.~\ref{sing_spec}) inconsistent with the
photometric classification as sdB+K7 by Allard et al. (\cite{alwe94}, where
all three stars seen in Fig.~\ref{resolv} were included in the measurements) 
but in accordance with the early spectroscopic classification by Green et al. 
(\cite{grsc86}).

\begin{figure}
\vspace{7.5cm}
\includegraphics{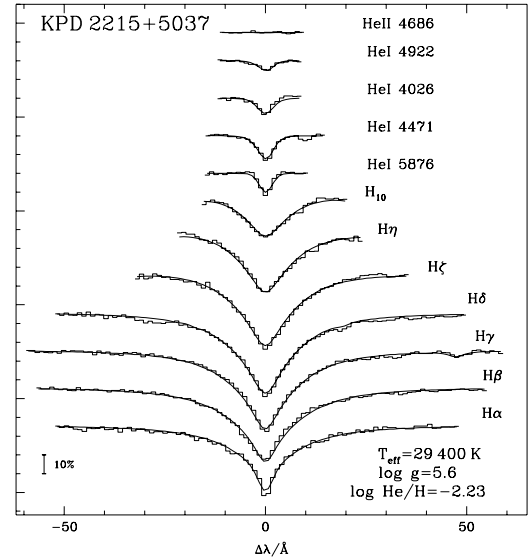}
\caption{Spectral fit for the sdB star
KPD~2215$+$5037. H$\epsilon$ is excluded from the fit because of 
contamination by interstellar \ion{Ca}{ii}.\label{kpd_fit}}
\end{figure}

We derived the atmospheric parameters \teff, \logg\ and helium abundance 
simultaneously for the single stars by
matching a grid of synthetic spectra derived from H and He line blanketed 
NLTE model atmospheres (Napiwotzki \cite{napi97}) to the data.
 For temperatures 
below 27,000\,K we used the metal line blanketed LTE model atmospheres of 
Heber et al. (\cite{here00}). 
The synthetic spectra were convolved beforehand with 
a Gaussian profile of the appropriate FWHM to account for the instrumental 
profile.
Results are given in Table~\ref{tab_spec} and Fig.~\ref{kpd_fit} displays 
the fit for KPD~2215$+$5037 as an example.

\begin{figure}[!b]
\vspace{7.5cm}
\includegraphics{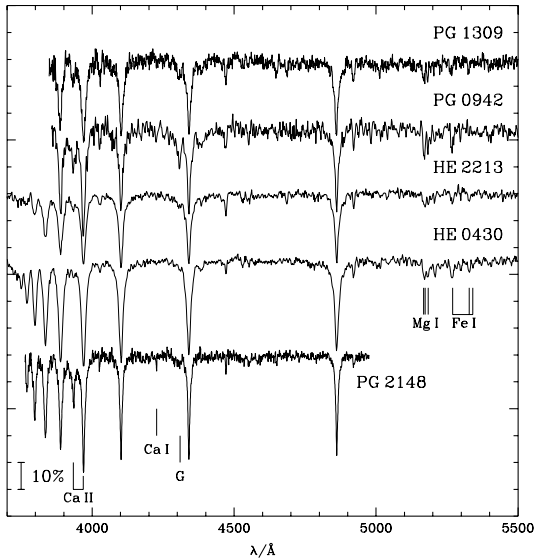}
\caption{Comparison of normalized spectra of four programme stars 
to PG~1309$-$078, which is already known to be a spectroscopic binary
containing an sdB.
The spectral features indicative of a cool companion are marked.
\label{bin_spec}}
\end{figure} 

\section{Photometric data for our programme stars
 \label{app_phot}}

In Tables~\ref{medium} and \ref{broad} we compile the photometric 
data collected from literature and used in the photometric deconvolution.

\begin{table*}[!t]
\caption[]{Str\"omgren photometry and UV spectrophotometry
for our programme stars. Str\"omgren photometry is taken 
from Green (\cite{gree80}, G80), Kilkenny (\cite{kilk84}, K84;  
\cite{kilk87}, K87), Moehler et al. (\cite{mori90}, M90), 
Theissen et al. (\cite{thmo93}, T93), Wesemael et al. (\cite{wefo92}, W92).
The IUE data were obtained from the IUE final archive 
(http://archive.stsci.edu/iue/). \label{medium}}
\begin{tabular}{l|llllll|lllll}
\hline
Star & $y$ & $b-y$ & $u-b$ & $m_1$ & $c_1$ & Ref. & \multicolumn{2}{c}{IUE} \\ 
     &     &       &       &       &       &      &  SWP & LWP \\
\hline
PB~6107 & \magpt{12}{897} & \magpt{+0}{032} & \magpt{+0}{112} & \magpt{+0}{052} 
 & & W92 & & \\
 & \magpt{12}{889} & \magpt{+0}{026} & & \magpt{+0}{092} & \magpt{-0}{094} 
 & M90 & & \\ 
 & \magpt{12}{89} & \magpt{+0}{01} & \magpt{+0}{10} & \magpt{+0}{05} & & G80
 & & \\ 
 & \magpt{12}{907} & \magpt{+0}{018} & & \magpt{+0}{093} & \magpt{-0}{109} 
 & K87 & & \\
PG~0105$+$276 & \magpt{14}{481} & \magpt{+0}{022} & \magpt{-0}{194} & 
 \magpt{+0}{023} & & W92 & 56271 & \\
PHL~1079 & \magpt{13}{278} & \magpt{+0}{003} & \magpt{+0}{106} & 
 & \magpt{-0}{109} & K84 & 42338 &  21098 \\
PG~0749$+$658 & \magpt{12}{135} & \magpt{-0}{032} & \magpt{+0}{131} & 
 \magpt{+0}{087} & & W92 & & \\
TON~1281 & \magpt{13}{371} & \magpt{+0}{094} & \magpt{+0}{175} & 
 \magpt{+0}{065} & & W92 & 56384 & \\
TON~139 & \magpt{12}{796} & \magpt{+0}{111} & \magpt{+0}{364} & 
 \magpt{+0}{055} & & W92 & & \\
PG~1309$-$078 & \magpt{14}{11} & \magpt{+0}{07} & \magpt{+0}{06} & 
 \magpt{+0}{18} & & G80 & & \\
PG~1449$+$653 & \magpt{13}{580} & \magpt{+0}{041} & \magpt{+0}{047} 
 & \magpt{+0}{034} & & W92 & 34298 & \\
PG~1511$+$624 & \magpt{14}{421} & \magpt{+0}{049}  & \magpt{-0}{002} & 
 \magpt{+0}{005} & & W92 & 39370, 57359, 57361 & 18491\\
% & & & & & & & LWP 18491 \\
PG~1558$-$007 & \magpt{13}{528} & \magpt{-0}{011} & \magpt{+0}{244} & 
 \magpt{+0}{091} & & W92 & & \\
PG~1636$+$104 & \magpt{14}{090} & \magpt{+0}{169} & \magpt{+0}{426} & 
 \magpt{+0}{056} & & W92 & & \\ 
PG~1656$+$213 & & & & & & & 39422 & 18542 \\
TON~264 & \magpt{14}{070} & \magpt{+0}{008} & \magpt{-0}{053} & 
 \magpt{+0}{070} & & W92 & 39422 & 18542 \\
PG~1718$+$519 & \magpt{13}{686} & \magpt{+0}{102} & \magpt{+0}{307} & 
 \magpt{+0}{084} & & W92 & 41571 & 20308 \\
              & \magpt{13}{694} & \magpt{+0}{131} & & \magpt{+0}{094} & 
 \magpt{-0}{095} & T93 &  & \\
PG~2148$+$095 & \magpt{13}{037} & \magpt{+0}{028} & \magpt{+0}{087} & 
 \magpt{+0}{066} & & W92 & 56148 & \\
KPD~2215$+$5037 & \magpt{13}{739} & \magpt{-0}{026} & \magpt{+0}{034} & 
 \magpt{+0}{068} & & W92 & & \\
PG~2259$+$134 & \magpt{14}{478} & \magpt{-0}{038} &  & 
 \magpt{+0}{082} & \magpt{-0}{089} & M90 & 44821, 56182 & 23244 \\
PG~2259$+$134 & \magpt{14}{545} & \magpt{-0}{069} & \magpt{-0}{011} & 
 \magpt{+0}{088} & & W92 & &  \\
BD~$-7^\circ$5977 & & & & & & & 31030 & 10815 \\
\hline
\end{tabular}
\end{table*}

\begin{table*}
\caption[]{$BVRI$ (Allard et al. \cite{alwe94}), $UBVI$ 
(Ferguson et al. \cite{fegr84}), HST $R$ (this paper)
and infrared broadband photometry (UT98: Ulla \& Thejll \cite{ulth98},
2MASS: 2MASS 2$^{\rm nd}$ incremental data release,
http://irsa.ipac.caltech.edu/applications/2MASS/BasicSearch/) 
for our programme stars\label{broad}}
\begin{tabular}{l|llll|l|llll}
\hline
Star & $V$ & $B-V$ & $V-R$ & $R-I$ & $R_{\rm HST}$ & $J$ & $H$ & $K$ & Ref. \\ 
\hline
PB~6107 & \magpt{12}{881} & \magpt{-0}{038} & \magpt{+0}{070} & 
 \magpt{+0}{096} & \magpt{12}{80} & & & \\
PG~0105$+$276 & \magpt{14}{448} & \magpt{-0}{087} &  \magpt{+0}{086} &
 \magpt{+0}{127} & \magpt{14}{36} & \magpt{14}{347} & \magpt{13}{821}  &
 \magpt{13}{721} & 2MASS \\ 
PHL~1079 & & & & & \magpt{13}{24} & \magpt{12}{55} & \magpt{12}{23} &
 \magpt{12}{04} & UT98\\ 
HE~0430$-$2457 & \magpt{14}{155}$^1$ & \magpt{-0}{046}$^1$ & 
 \magpt{+0}{085}$^1$ & & \magpt{14}{07} & \magpt{13}{619} & \magpt{13}{315}
 & \magpt{13}{208}  & 2MASS \\ 
PG~0749$+$658 & \magpt{12}{121} & \magpt{-0}{106} & \magpt{+0}{021} & 
 \magpt{+0}{072} & \magpt{12}{14} & & & & \\
PG~0942$+$461 & & & & & \magpt{13}{96} & \magpt{13}{612} & \magpt{13}{172} & 
 \magpt{13}{084} & 2MASS \\
TON~1281 & \magpt{13}{439} & \magpt{+0}{094} & 
 \magpt{+0}{156} & \magpt{+0}{176} & \magpt{13}{27} & \magpt{12}{758} & 
 \magpt{12}{503} & \magpt{12}{448} & 2MASS \\
TON~139 & & & & & \magpt{12}{65} & \magpt{12}{10} & \magpt{11}{92} & 
 \magpt{11}{93} & UT98\\
PG~1309$-$078 & & & & & \magpt{14}{05} & \magpt{13}{558} & \magpt{13}{259} & 
 \magpt{13}{162} & 2MASS \\
PG~1449$+$653 & \magpt{13}{611} & \magpt{-0}{035} & 
 \magpt{+0}{073} & \magpt{+0}{110} & \magpt{13}{57} & & & \\
PG~1511$+$624 & \magpt{14}{527} & \magpt{-0}{022} & \magpt{+0}{113} 
 & \magpt{+0}{142} & \magpt{14}{38} & \magpt{14}{114} & \magpt{13}{813} & 
 \magpt{13}{883} & 2MASS \\
PG~1558$-$007 & \magpt{13}{541} & \magpt{-0}{064} & \magpt{+0}{012} 
 & \magpt{+0}{110} & \magpt{13}{55} & & & & \\
PG~1601$+$145 & \magpt{14}{424} & \magpt{+0}{028} & \magpt{+0}{180} & 
 \magpt{+0}{347} & \magpt{14}{37} & \magpt{13}{918} & \magpt{13}{578} & 
 \magpt{13}{600} & 2MASS \\
PG~1636$+$104 & \magpt{14}{039} & \magpt{+0}{193} & \magpt{+0}{191} & 
 \magpt{+0}{196} & \magpt{13}{85} & & & & \\
TON~264 & \magpt{14}{074} & \magpt{-0}{083} & \magpt{+0}{066} & \magpt{+0}{136} 
 & \magpt{14}{02} & & & & \\
PG~1718$+$519 & \magpt{13}{733} & \magpt{+0}{113} & \magpt{+0}{156} & 
 \magpt{+0}{132} & \magpt{13}{53} & \magpt{13}{008} & \magpt{12}{716} 
 & \magpt{12}{664} & 2MASS \\
PG~2148$+$095 & \magpt{13}{021} & \magpt{-0}{024} & \magpt{+0}{060} 
 & \magpt{+0}{096} & \magpt{12}{98} & \magpt{12}{18} & \magpt{12}{34} & 
 \magpt{12}{06} & UT98 \\
HE~2213$-$2212 & & & & & \magpt{14}{01} & \magpt{13}{686} & \magpt{13}{292} & 
 \magpt{13}{236} & 2MASS\\
KPD~2215$+$5037 & \magpt{13}{664} & \magpt{-0}{093} & \magpt{+0}{015} 
 & \magpt{+0}{052} & \magpt{13}{91} & & & & \\
PG~2259$+$134 & & & & & \magpt{14}{70} & \magpt{15}{795} & \magpt{15}{220} & 
 \magpt{14}{523} & 2MASS\\
BD~$-7^\circ$5977 & \magpt{10}{55}$^2$ & \magpt{+0}{51}$^2$ & & & \magpt{10}{05} 
 & \magpt{8}{97} & \magpt{8}{48} & \magpt{8}{38}& UT98\\
 & & & & & & \magpt{9}{017} & \magpt{8}{526} & \magpt{8}{448} & 2MASS \\
\hline
Star & $V$ & $B-V$ & $U-B$ & $V-I$ & $R_{\rm HST}$ & $J$ & $H$ & $K$ & Ref. \\ 
\hline
PG~1421$+$345 & \magpt{14}{78} & \magpt{-0}{14} & \magpt{-0}{89} & 
\magpt{+0}{63} & \magpt{14}{59} & \magpt{14}{035} & \magpt{13}{716} & 
 \magpt{13}{678} & 2MASS \\
PG~1601$+$145 & \magpt{14}{50} & \magpt{+0}{01} & \magpt{-0}{92} & 
\magpt{+0}{41} & & & & & \\ 
PG~1656$+$213 & \magpt{14}{88} & \magpt{-0}{20} & \magpt{-0}{73} & & 
\magpt{14}{73} & & & & \\
\hline
\end{tabular}
\begin{tabular}{l}
$^1$ Altmann (priv. comm.)\\
$^2$ Derived from Tycho photometry ($V_T$ = \magpt{10}{6}, $(B-V)_T$ = 
\magpt{+0}{6}) using the transformation given in Perryman (\cite{perr97}).\\
\end{tabular}
\end{table*}
\end{document}